\documentclass[
  journal=pasa,
  manuscript=research-paper,
  year=2026,
  volume=YY,
]{cup-journal}

\usepackage{multirow}
\usepackage{booktabs}
\usepackage{graphicx}
\usepackage{subcaption}
\usepackage{amsmath}
\usepackage{amssymb}
\usepackage{siunitx}
\usepackage{tabularx}
\usepackage{rotating} 

\newcommand{\sigmaRM}{$\sigma_{\rm RM}$}

\newcommand{\hmshh}{\ensuremath{^\mathrm{h}}}
\newcommand{\hmsmm}{\ensuremath{^\mathrm{m}}}
\newcommand{\hmsss}{\ensuremath{^\mathrm{s}}}
\newcommand{\degree}{\ensuremath{^{\circ}}}
\newcommand{\arcm}{\ensuremath{^{\prime}}}
\newcommand{\arcs}{\ensuremath{^{\prime\prime}}}

\newcommand{\frba}{FRB 20190611B}
\newcommand{\frbb}{FRB 20210407E}
\newcommand{\frb}{FRB 20250607A}

\newcommand{\DM}{\ensuremath{\rm DM}}
\newcommand{\pccm}{\,\ensuremath{\rm pc\,cm^{-3}}} 
\newcommand{\DDM}{$\Delta$\DM}
\newcommand{\RM}{\ensuremath{\rm RM}}
\newcommand{\radm}{\,\ensuremath{\rm rad\,m^{-2}}}

\title{FRB 20250607A: Weirder Than Fiction}

\author{J.~C.~F.~Balzan}
\affiliation{International Centre for Radio Astronomy Research (ICRAR), Curtin University, Bentley, WA 6012, Australia}
\email[J.~C.~F.~Balzan]{joel.balzan@icrar.org}

\author{A.~Bera}
\affiliation{ASTRON, Netherlands Institute for Radio Astronomy, Postbus 2, 7990 AA Dwingeloo, Netherlands}
\alsoaffiliation{International Centre for Radio Astronomy Research (ICRAR), Curtin University, Bentley, WA 6012, Australia}

\author{C.~W.~James}
\affiliation{International Centre for Radio Astronomy Research (ICRAR), Curtin University, Bentley, WA 6012, Australia}

\author{B.~W.~Meyers}
\affiliation{Australian SKA Regional Centre (AusSRC), Curtin University, Kent Street, Bentley, WA 6102, Australia}
\alsoaffiliation{International Centre for Radio Astronomy Research (ICRAR), Curtin University, Bentley, WA 6012, Australia}

\author{A.~T.~Deller}
\affiliation{Centre for Astrophysics and Supercomputing, Swinburne University of Technology, Hawthorn, VIC, 3122, Australia}
\alsoaffiliation{ARC Centre of Excellence for Gravitational Wave Discovery (OzGrav), Post Office Box 218, Hawthorn, VIC 3122, Australia}

\author{R.~M.~Shannon}
\affiliation{Centre for Astrophysics and Supercomputing, Swinburne University of Technology, Hawthorn, VIC, 3122, Australia}

\author{Z.~Wang}
\affiliation{International Centre for Radio Astronomy Research (ICRAR), Curtin University, Bentley, WA 6012, Australia}

\received{dd Mmm YYYY}
\revised{dd Mmm YYYY}
\accepted{dd Mmm YYYY}
\published{22 September 202X}


\begin{document}

\begin{abstract}
We present a detailed analysis of \frb, a two-component FRB detected by the CRAFT survey which exhibits several unusual component-dependent properties, including apparent differential \DM, circular polarisation, and time-resolved \RM, together with a polarisation-angle (PA) swing that persists through the scattering tail. The scattering timescale and scintillation bandwidth are inconsistent with a single thin-screen origin, placing the dominant scattering screen within $\lesssim61$~kpc of the FRB source. The lack of PA flattening in the scattering tail is a behaviour not previously reported in an FRB or pulsar and difficult to reconcile with a simple propagation origin. We find no statistically significant differential \DM\ between the two components despite an apparent, and visually obvious, difference in their dispersion, indicating that current uncertainties may be overly conservative. If this apparent differential \DM\ is true, it implies a substantial difference in electron column density between the component sightlines. Two further CRAFT FRBs, \frba\ and \frbb, show similarly large component-dependent electron-column differences, corresponding to electron-density contrasts consistent with plasma local to a neutron-star magnetosphere. We find no evidence for generalised Faraday rotation, and \frbb\ shows component-dependent PA structure. Taken together, these results suggest that the observed \DM, circular polarisation, and PA behaviour may arise from differing sightlines through an inhomogeneous or evolving magnetosphere rather than from independent phenomena. Multi-component FRBs therefore provide a sensitive probe of plasma structure local to their progenitors.
\end{abstract}

\maketitle

\section{Introduction}\label{sec: intro}

Fast radio bursts (FRBs) are intense and typically highly polarised flashes of radio light of cosmological origin. As FRBs propagate through plasma, they undergo a variety of propagation effects, including dispersion, scattering, and Faraday rotation, making them excellent probes of the interstellar and intergalactic media (IGM). Dispersion is caused by travelling through cold, ionised media and imparts a frequency-dependent delay of arrival time which can be quantified by the dispersion measure,
\begin{equation}\label{eq: dm}
    {\rm DM} = \int_0^d \frac{n_e(l)}{1+z}\, {\rm d}l\; {\rm pc\,cm^{-3}},
\end{equation}
where $n_e$ is the electron density along the line-of-sight (LOS), $z$ is the redshift, and $d$ is the path length. The \DM\ is often used to estimate the distance to FRBs using the Macquart ($z$-\DM) relation~\citep[][]{macquart2020}; there are many contributions to the \DM\ of an FRB, including the Milky Way, IGM, host galaxy of the FRB, and the local environment of the FRB progenitor. Separating these contributions is difficult; however, the \DM\ can still be used to infer large-scale properties of the IGM \citep[e.g.,][]{li2026b}, and smaller-scale local environment evolution \citep[e.g.,][]{2021Natur.598..267L,2022Natur.609..685X,2023MNRAS.526.3652K,niu2026}.

In order to investigate FRBs at high time resolution (HTR), the observed signal must be dedispersed using an estimate of its \DM. Accurate dedispersion is essential because the inferred burst morphology, temporal structure, polarisation properties, and scattering measurements critically depends on the accuracy of dedispersion. FRBs often exhibit complex spectro-temporal structure and variability \citep[e.g.,][]{hessels2019,kumar2023a}, making it difficult to disentangle intrinsic burst structure from dispersion~\citep[e.g.,][]{feng2026}. An incorrectly estimated \DM\ therefore risks not only misrepresenting the burst's temporal structure, but also biasing the polarimetric and scattering diagnostics used to probe the FRB's local environment and propagation path \citep[e.g.,][]{wang2026a}.

In several individual FRBs, sub-components within the same burst appear to favour different optimal \DM{s} \citep[e.g.,][]{2020ApJ...891L..38C,platts2021,2023MNRAS.526.2039H,2023MNRAS.526.3652K,faber2024,2026ApJ...998..276Z,2026MNRAS.546ag090O}. If genuine, such intra-burst \DM\ differences imply that successive sub-components --- separated by only milliseconds --- encounter different integrated electron columns along the LOS, potentially providing a probe of plasma structure on extremely small spatial scales near the source. However, apparent \DM\ differences may also arise from plasma lensing, scattering, intrinsic burst morphology, or biases introduced by the optimisation method itself.


In this work we present a detailed propagation and polarimetric analysis of \frb, a multi-component FRB exhibiting apparent component-dependent \DM, distinct scattering and scintillation screens consistent with a two-screen geometry, time-resolved rotation measure variations, and a polarisation-angle swing that persists through its scattering tail despite the flattening expected from a simple thin-screen model. We use two additional CRAFT FRBs, \frba\ and \frbb\ \citep{2025arXiv250517497S}, for comparison of component-\DM\ and electron-density-contrast behaviour, with \frbb\ also showing component-dependent PA structure. We discuss possible physical origins for these combined signatures, including plasma lensing, magnetospheric propagation effects, and a two-screen scattering geometry, and consider whether they may share a common origin.

\section{\frb}
\label{sec:Methods and Results}

\subsection{Detection, localisation, and pre-processing}
\label{sec:Detection and Localisation}

On June 7, 2025, at UTC 02:17:06 (MJD = 60833.10994587), the Australian Square Kilometre Array Pathfinder (ASKAP) Commensal Real-time ASKAP Fast Transients (CRAFT) incoherent-sum (ICS) \citep{2019Sci...365..565B,2025PASA...42...36S} and Coherent CRAFT Upgrade \citep[CRACO;][]{2025PASA...42....5W} FRB detection systems reported the discovery of \frb\ at a centre frequency of 919.5~MHz with a bandwidth of 336~MHz. The CRACO detection reported a signal-to-noise ratio of 75 and a \DM\ of 336.0~\pccm.

A download of the full 12.4~s of 1-bit voltage data was triggered by the real-time system and transferred to the Ngarrgu Tinderbeek supercomputer at Swinburne University of Technology. Calibration observations of the polarisation calibrator PSR J0835$-$4510 (Vela) and the bandpass calibrator PKS B0407$-$658 were obtained 4.74 and 4.94~h, respectively, after the initial FRB detection, each comprising 12.4~s of 1-bit voltage data. The raw voltages of \frb\ were processed and the calibrator observations were applied using the CRAFT Effortless Localisation and Enhanced Burst Inspection (CELEBI) pipeline~\citep{2023A&C....4400724S,2026arXiv260506766G}. This yielded a sub-arcsecond localisation of RA = 02\hmshh 32\hmsmm 43.445 \hmsss\,$\pm$\,0.417\arcs, Dec = $-$39\degree 20\arcm 27.949\,$\pm$\,0.414\arcs, with a $1\sigma$ uncertainty ellipse of major axis 0.418\arcs, minor axis 0.413\arcs, and position angle $-69.64$\degree. This uncertainty is wholly dominated by the systematic uncertainty of the Rapid ASKAP Continuum Survey (RACS1-low) reference catalogue~\citep[][]{jaini2025} used to astrometrically register the FRB position, as the statistical precision of this (very bright) burst is just 0.049\arcs\ in RA and 0.038\arcs\ in Dec. The CELEBI pipeline also measured a structure-maximised \DM\ of $335.87^{+0.34}_{-0.26}$~\pccm\ using the method of \citet{2023ApJ...954...37S}, implemented in the SHRINE package\footnote{\url{https://github.com/marcinglowacki/SHRINE}}. \frb\ has been localised to a host galaxy at a redshift of $z = 0.2106$ (Gordan et al., in prep.).

The calibrated data products are further processed using the \texttt{ILEX}\footnote{\url{https://github.com/tdial2000/ILEX}} software package, which includes the following steps: baseline subtraction, Stokes dynamic spectra (Figure~\ref{fig:htr_plot3}) formation with $\delta t = 0.016688$\,ms and $\delta\nu = 0.120$\,MHz time and frequency resolutions, respectively, linear polarisation debiasing~\citep{2001ApJ...553..341E} and rotation measure (\RM) correction~\citep{2026ApJS..283...28V}. 

\begin{figure}[ht!]
    \centering
    \includegraphics[width=\linewidth]{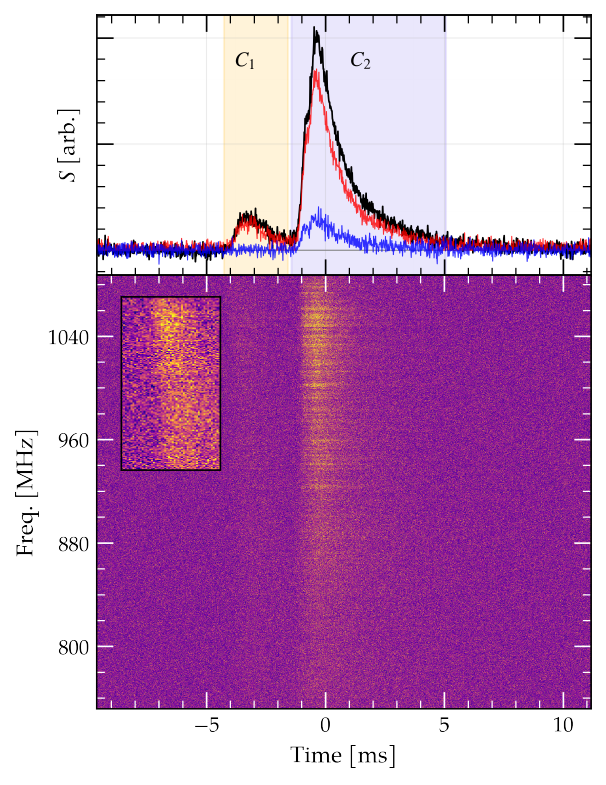}
 \caption{
    \frb\ dedispersed at \DM\,=\,$335.87$~\pccm\ and derotated at $\RM\,=-428.2$~\radm with time resolution $\delta t = 0.016688$\,ms and frequency resolution $\delta\nu = 0.120$\,MHz. The bottom panel shows the time-frequency dynamic spectrum, and the top shows the frequency-summed pulse profile, with total intensity in black, linear polarisation in red, and circular polarisation in blue. The yellow and purple shaded regions, labelled $C_1$ and $C_2$, show the time crops used for the individual component analysis, indicating component one and component two, respectively. The dynamic spectrum additionally shows an inset of $C_1$ decimated by a factor of four in time and 16 in frequency. 
    }
\end{figure}\label{fig:htr_plot3}

\subsection{DM optimisation}\label{subsec: shrine dm}
Figure~\ref{fig:htr_plot3} shows \frb\ at the CELEBI dispersion measure, \DM$_{\rm C}=335.87^{+0.34}_{-0.26}$~\pccm, obtained by SHRINE structure maximisation on a global burst envelope. While this value maximises the overall burst structure, both temporally separated emission components, by eye, show residual frequency drifts, indicating potential differential \DM\ between components. To investigate this, we perform \DM\ optimisation independently on each emission component. For each case, fixed time windows are manually selected from the CELEBI dedispersed Stokes $I$ time series (the yellow and purple shaded regions in Figure~\ref{fig:htr_plot3}). All subsequent component-dependent analyses are performed independently within these windows.


We use SHRINE to determine the \DM$_{\rm struct}$ that maximises structure in the dedispersed time series. For each component we construct a trial \DM\ grid centred on the CELEBI value, spanning the range listed in Table~\ref{tab:dm_results}, and incoherently dedisperse each trial relative to the maximum observing frequency. Samples shifted outside the observing window are zero-padded. For each trial \DM\ we compute the frequency-summed Stokes $I(t)$, constructing $I(t,{\rm DM})$. The structure parameter is defined as the norm of the derivative of the smoothed profile, $S({\rm DM}) = ||dI'/dt||$, where $I'(t,{\rm DM})$ is the smoothed data obtained using the discrete cosine transform (DCT)-based filtering procedure described by \citet{2023ApJ...954...37S}. The smoothing scale is determined by the spectral cutoff $k_c$, which defines the maximum DCT frequency retained in the low-pass filter. The optimal \DM$_{\rm struct}$ is taken to be the value that maximises $S({\rm DM})$. Uncertainties on \DM$_{\rm struct}$ are calculated following \citet{2023ApJ...954...37S}, which accounts for correlations between nearby trial \DM{s}. 

\begin{table*}[ht!]
    \centering
    \caption{SHRINE Structure maximisation \DM\ optimisation results for burst components of \frba, \frbb, and \frb, where \DM$_{\rm C}$ is the original CELEBI \DM.}
    \label{tab:dm_results}
    \footnotesize
    \setlength{\tabcolsep}{5pt}
    \begin{tabular}{@{}l r c r l r r@{}}
        \toprule
        FRB & \multicolumn{1}{c}{\DM$_{\rm C}$} & \multicolumn{1}{c}{\DM\ Range} & \multicolumn{1}{c}{Step}
            & Component
            & \multicolumn{1}{c}{Optimised \DM}
            & \multicolumn{1}{c}{$\Delta n_e$} \\
            & \multicolumn{1}{c}{($\mathrm{pc\,cm^{-3}}$)} & \multicolumn{1}{c}{($\mathrm{pc\,cm^{-3}}$)} & \multicolumn{1}{c}{($\mathrm{pc\,cm^{-3}}$)}
            & & \multicolumn{1}{c}{($\mathrm{pc\,cm^{-3}}$)} & \multicolumn{1}{c}{($\mathrm{cm^{-3}}$)} \\
        \midrule
        \multirow[t]{6}{*}{\frba} & \multirow[t]{6}{*}{$322.65^{+0.14}_{-0.08}$} & \multirow[t]{6}{*}{$[322.0, 323.3]$} & \multirow[t]{6}{*}{$0.003$}
            & $C_1$   & $322.642^{+0.054}_{-0.060}$        & \multicolumn{1}{c}{} \\
        \cmidrule(l){5-7}
        &   &   &   & $C_2$   & $322.735^{+0.261}_{-0.093}$        & \multicolumn{1}{c}{} \\
        \cmidrule(l){5-7}
        &   &   &   & $\Delta\mathrm{DM}_{2-1}$ & $0.103^{+0.269}_{-0.116}$   & $(1.08^{+2.82}_{-1.21})\times10^{10}$ \\
        \midrule
        \multirow[t]{6}{*}{\frbb} & \multirow[t]{6}{*}{$1784.82^{+0.17}_{-0.09}$} & \multirow[t]{6}{*}{$[1784.32, 1785.32]$} & \multirow[t]{6}{*}{$0.0007$}
            & $C_1$   & $1784.771^{+0.087}_{-0.064}$       & \multicolumn{1}{c}{} \\
        \cmidrule(l){5-7}
        &   &   &   & $C_2$   & $1784.921^{+0.082}_{-0.067}$       & \multicolumn{1}{c}{} \\
        \cmidrule(l){5-7}
        &   &   &   & $\Delta\mathrm{DM}_{2-1}$ & $0.148^{+0.105}_{-0.110}$   & $(2.76^{+1.96}_{-2.04})\times10^{10}$ \\
        \midrule
        \multirow[t]{6}{*}{\frb} & \multirow[t]{6}{*}{$335.87^{+0.34}_{-0.26}$} & \multirow[t]{6}{*}{$[334.87, 336.87]$} & \multirow[t]{6}{*}{$0.004$}
            & $C_1$   & $335.942^{+0.144}_{-0.100}$        & \multicolumn{1}{c}{} \\
        \cmidrule(l){5-7}
        &   &   &   & $C_2$   & $335.830^{+0.228}_{-0.120}$        & \multicolumn{1}{c}{} \\
        \cmidrule(l){5-7}
        &   &   &   & $\Delta\mathrm{DM}_{2-1}$ & $-0.109^{+0.253}_{-0.195}$  & $(-4.55^{+10.56}_{-8.12})\times10^{9}$ \\
        \bottomrule
    \end{tabular}
\end{table*}

Applying the structure-maximisation metric to the two components of \frb\ yields a difference of $\Delta\mathrm{DM}_{2-1}=-0.109^{+0.253}_{-0.195}$~\pccm. Within their uncertainties, both component \DM{s} (see Table~\ref{tab:dm_results}) are consistent with the original CELEBI value, indicating no statistically significant evidence for differential dispersion between the two emission components. We tested with larger time resolutions ($2,4,8\times\delta t$) and found no meaningful improvement in constraining the error bounds. 

Incidentally, in the CRAFT HTR sample we find two other FRBs that show apparent differential \DM\ between components, \frba\ and \frbb~\citep[][]{2025arXiv250517497S}. We repeat the same analysis, obtaining component \DM{s} that are likewise consistent with their respective CELEBI dispersion measures within the uncertainties. The complete optimisation results are summarised in Table~\ref{tab:dm_results}, while the corresponding structure-maximisation scans and dedispersed burst morphologies are shown in Figures~\ref{fig:190611_dm_structure}, \ref{fig:210407_dm_structure}, and \ref{fig:250607_dm_structure}.

\subsubsection{SHRINE DM uncertainties}\label{subsec: shrine dm uncertainties}
\begin{figure*}
    \centering
    \includegraphics[width=\linewidth]{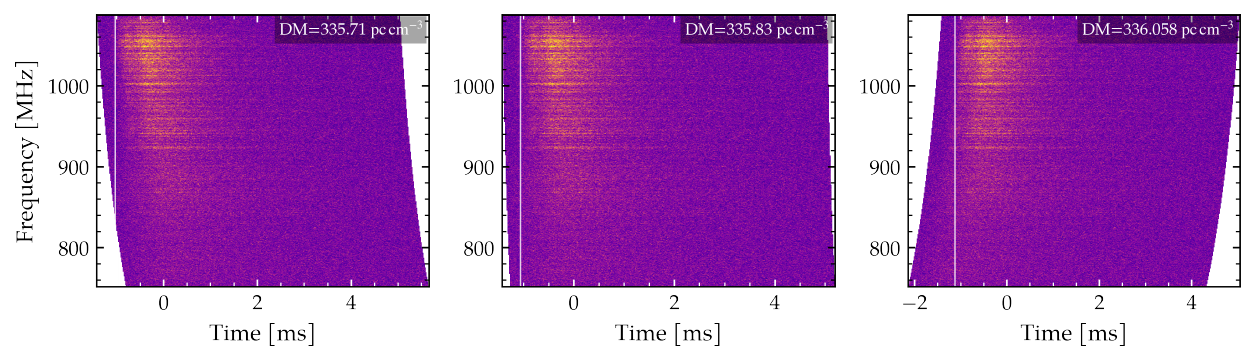}
    \caption{Dedispersed dynamic spectra of \frb\ component 2 evaluated at the lower bound (left), optimum (centre), and upper bound (right) \DM\ values from the SHRINE structure-maximisation confidence interval. The vertical white line indicates the onset of the pulse at the maximum frequency.}
    \label{fig: DM error range}
\end{figure*}

Despite finding no statistical evidence for different \DM{s}, visual inspection suggests that the burst components may exhibit genuine \DM\ differences. Furthermore, dispersing the bursts to the $\pm1\sigma$ limits of the uncertainty ranges derived from \DM$_{\rm struct}$ produces increasingly evident residual frequency-dependent structure (Figure~\ref{fig: DM error range}), suggesting that these uncertainties may be overly conservative. We inspected both the implementation described by \citet{2023ApJ...954...37S} and the corresponding SHRINE codebase, but found no obvious issues that would explain this behaviour.

We therefore suspect that the inflated uncertainties may arise, at least in part, from limitations of the original filter design used in the structure metric. The low-pass filter is described by the spectral response,
\begin{equation}\label{eq:lpf}
    f(k) = \frac{1}{1+\big(\frac{k}{k_c}\big)^{2O}},
\end{equation}
where $k=1,2,\dots,k_c$, and $O$ is the order of the filter. Although \citet{2023ApJ...954...37S} empirically tune the spectral cutoff, $k_c$, the filter order ($O=3$) and Equation~\eqref{eq:lpf} are explored only qualitatively and are not compared against alternative filter choices. It is therefore possible that this fixed filter order and shape do not generalise optimally across the diverse range of temporal structures present in real FRB data, leading to conservative uncertainty estimates. A systematic exploration of alternative filter forms and their optimisation alongside $k_c$ may therefore be required to improve the accuracy of the uncertainty estimates. We defer such an investigation to a separate work (Balzan et al., in prep.). Consequently, while the fitted \DM\ values remain formally consistent with the CELEBI \DM, low-level differential \DM\ between burst components cannot be ruled out.

\subsection{\frb\ scintillation and scattering}\label{sec:Scintillation Bandwidth}
\begin{figure}[h!]
    \centering

    \begin{subfigure}{\linewidth}
        \centering
        \includegraphics[width=\linewidth]{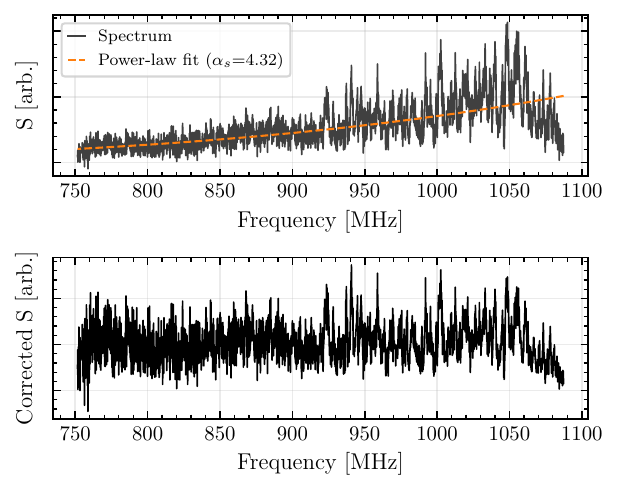}
        \caption{Top panel: \frb\ $C_2$ time-averaged spectrum and power-law fit. Bottom panel: power-law-corrected spectrum according to Equation~\eqref{eq:residual_spectrum}.}
        \label{fig:scint_spectrum}
    \end{subfigure}

    \vspace{0.5em}

    \begin{subfigure}{\linewidth}
        \centering
        \includegraphics[width=\linewidth]{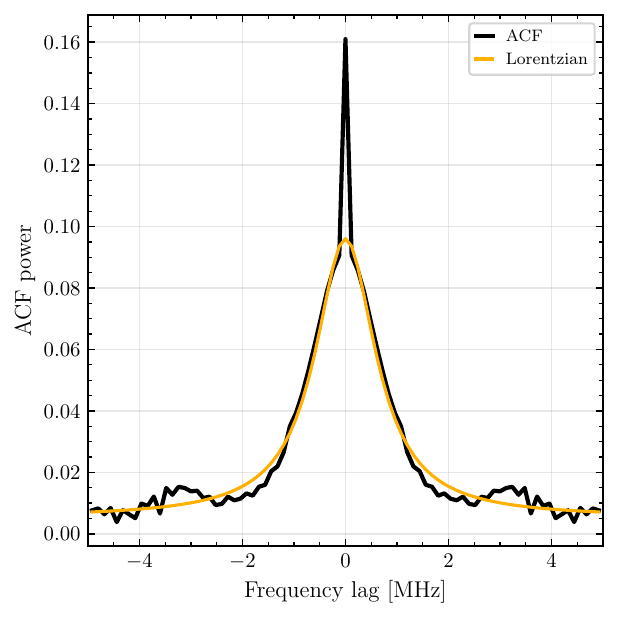}
        \caption{Frequency autocorrelation function of \frb\ $C_2$ spectrum (black), fitted with a single-component Lorentzian model (yellow) over frequency lags of 0--5 MHz. The best-fit decorrelation bandwidth is $\Delta \nu_{\rm d}=0.7113\pm0.031$ MHz, with modulation index $m_g = 0.3014$.}
        \label{fig:acf}
    \end{subfigure}
    \caption{Scintillation analysis of \frb\ $C_2$. (a) Spectrum with the fitted power-law bandpass model. (b) Frequency autocorrelation function and best-fitting Lorentzian model used to determine the scintillation decorrelation bandwidth.}
    \label{fig:scintillation}
\end{figure}

\begin{figure}[h!]
    \centering
    \includegraphics[width=1\linewidth]{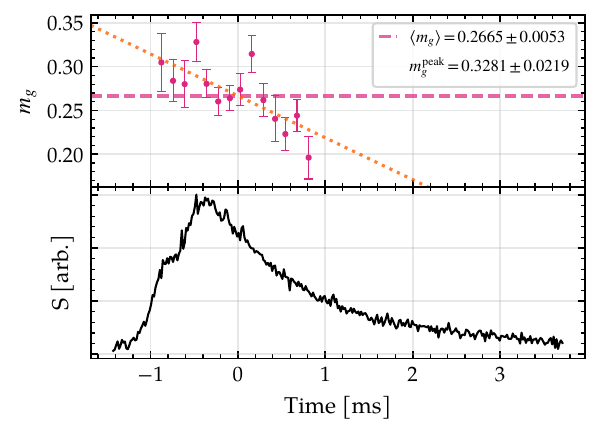}
    \caption{Time-resolved modulation index of \frb\ $C_2$, computed from the time-averaged spectrum in 0.09~ms bins. The horizontal-dashed line indicates the weighted mean, and the orange dotted line is the weighted linear least-squares fit ($\chi^2_\mathrm{red}=1.3$).}\label{fig: mod idx}
\end{figure}

\frb\ seems to exhibit both scintillation and scattering. We measured the scintillation bandwidth from \frb\ $C_2$ Stokes $I$ dynamic spectrum. The dynamic spectrum was cropped to the FWHM of the pulse profile, and time-averaged to form the burst spectrum (Figure~\ref{fig:scint_spectrum}). To remove the intrinsic spectral slope prior to scintillation analysis, we followed the approach of \citet{Macquart_2019}, modelling the mean burst spectrum as a power law of the form $S(\nu) \propto \nu^{\alpha_s}$ and fitting the spectral index using a least-squares fit in log-log space after smoothing the spectrum into 64 frequency bins. The best-fit spectral index was $\alpha_s = 4.317 \pm 0.269$. The fitted model was used to construct a smooth mean spectrum, which was then subtracted and normalised to form the fractional residual spectrum,
\begin{equation}\label{eq:residual_spectrum}
    \delta S(\nu) = \frac{S(\nu) - \bar{S}(\nu)}{\bar{S}(\nu)}.
\end{equation}
The residual spectrum was restricted to 760--1080~MHz for autocorrelation function (ACF) analysis to exclude any band-edge effects. We computed the frequency ACF of the residual spectrum and fit Lorentzian models after masking the zero lag noise spike, comparing 1-, 2-, and 3-component models out to 100~MHz maximum lag. We find that the ACF is best described by a single Lorentzian, yielding a scintillation bandwidth of $\Delta\nu_{\rm d} = 0.7113 \pm 0.031$~MHz (Figure~\ref{fig:acf}) at the centre frequency (919.5~MHz). The modulation index, $m_g = 0.3014 \pm 0.004$, was obtained from the square-root of the Lorentzian amplitude, and is consistent with the peak time-resolved modulation index, $m_g = 0.33 \pm 0.02$ (Figure~\ref{fig: mod idx}). We fit a weighted linear model with least-squares to the time-resolved $m_g$, weighting each point by the inverse variance of its measurement, and find a slope $b = -0.0479 \pm 0.0117$ms$^{-1}$ ($\chi^2_\mathrm{red} = 1.30$). The decreasing modulation index is expected if the Milky Way scattering screen partially resolves the scattered image of the source~\citep[][]{2023MNRAS.525.5653S,pradeepe.t.2025a}.

Due to limited bandwidth and spectral S/N, we were unable to reliably fit for the scintillation bandwidth index. We rule out self-noise as the origin of the observed frequency structures, as the reciprocal of our 0.016688~ms time resolution is 0.06~MHz, which is less than both our frequency resolution and the observed frequency structures. We discuss this further in Section~\ref{sec:two screen}.

\begin{figure*}[ht]
    \centering
    \includegraphics[width=\linewidth]{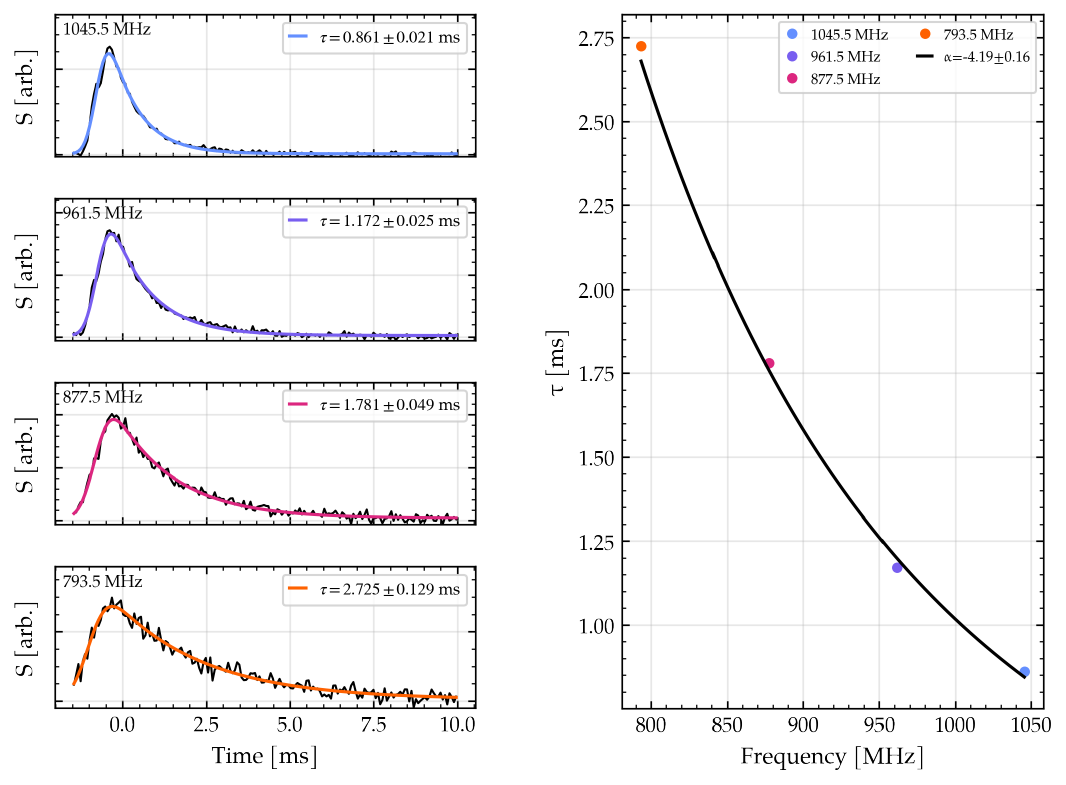}
    \caption{\frb\ component 2 (decimated in time by a factor of four) scattering-index fitting in four equal-bandwidth subbands spanning the observing band.}
    \label{fig: sc idx}
\end{figure*}

The scattering properties of \frb\ were characterised by fitting a scattered Gaussian pulse profile using least-squares across four subbands (Figure~\ref{fig: sc idx}). The frequency dependence was then fit using weighted least squares, with the uncertainties from the individual subband fits used as weights. This yielded a scattering index of $\alpha = -4.19 \pm 0.16$, consistent with both Gaussian and Kolmogorov turbulence. Using the best-fitting index, the scattering timescale measured from the full pulse profile at the band-centre frequency of 919.5~MHz is $\tau = 1.25 \pm 0.02$ ms. These values of $\tau$ and $\alpha$ were adopted for subsequent analysis and simulations.

Our measured scintillation bandwidth and scattering timescale violate the Fourier uncertainty relation $2\pi\tau\Delta\nu_{\rm d} \approx 1$, implying that the scattering and scintillation do not arise from a single thin scattering screen. The corresponding comparison with Galactic scattering predictions from NE2025 is presented in \ref{app:NE2025} and predicts $\tau^{\rm NE2025} = 7.67 \times 10^{-5}$\,ms and $\Delta \nu_{\rm d}^{\rm NE2025} = 2.41$\,MHz. The measured and NE2025-predicted scintillation bandwidths are broadly consistent in magnitude, with the prediction exceeding the measured value by a factor of $\sim3.4$, supporting that the Galactic screen can plausibly account for the observed diffractive scintillation. We discuss the two-screen interpretation further in Section~\ref{sec:two screen}.

\subsection{RM and depolarisation}\label{subsec: RM}
\begin{figure}[ht!]
    \centering
    \includegraphics[width=\linewidth]{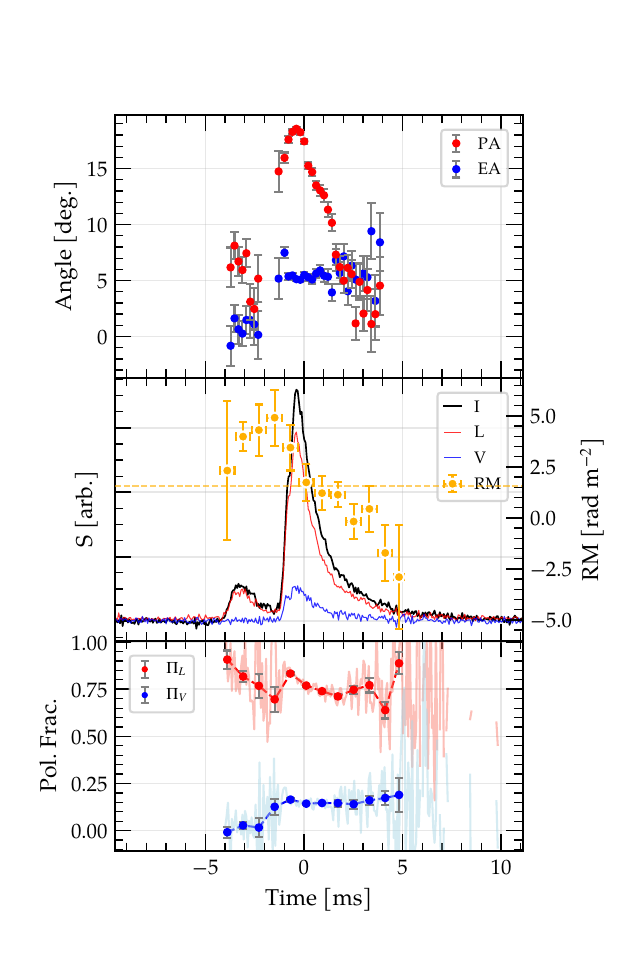}
    \caption{Time-resolved \RM\ and polarisation for \frb. The top panel shows the polarisation angle and ellipticity angle (decimated by a factor of 12). The middle panel presents Stokes $I$ (with $L$ and $V$ in red and blue, respectively; decimated by a factor of four) together with \RM\ estimates as a function of time in 12 bins; vertical error bars show the uncertainty in \RM, horizontal error bars indicate the time window over which the \RM\ was computed, and the horizontal dashed line is the mean \RM. The bottom panel shows the fractional linear and circular polarisation, $\Pi_L$ and $\Pi_V$. The middle and bottom panels are decimated in time by a factor of four.}\label{fig: RM}
\end{figure}

We derotate the average on-pulse spectra of \frb\ at $\RM\,=-428.2$~\radm with \texttt{RM-Tools}~\citep{2026ApJS..283...28V}. We then search for evidence of time-resolved \RM\ by selecting regions containing the burst emission and perform \RM\ synthesis in equal time bins (Figure~\ref{fig: RM}). Bins are excluded if the Stokes~$I$ S/N\,$<2$, or if the Faraday spectrum S/N\,$<5$. \frb\ shows clear residual time-resolved \RM\ structure after the initial de-rotation, which has been observed in other FRBs \citep[e.g.,][]{2024ApJ...969L..29B,2025ApJ...982..119B}, and pulsars \citep[e.g.,][]{noutsos2009,karastergiou2009,ilie2019}.


\begin{figure}[ht!]
    \centering
    \includegraphics[width=1\linewidth]{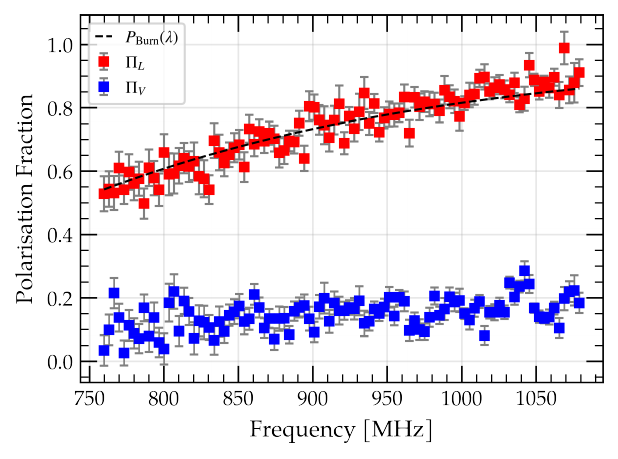}
    \caption{Burn's-law fit for $C_2$ of \frb\ with 100 frequency bins. Measured $\Pi_L$ (red squares) and $\Pi_V$ (blue squares) are shown as a function of frequency, with fitting performed via least-squares in $\lambda^2$ space excluding two bins on each band edge. \sigmaRM\,$ = 3.56 \pm 0.03$\radm.}
    \label{fig: burns law main}
\end{figure}

For the frequency-resolved analysis of \frb, we consider the high S/N second component. The selected on-pulse region was averaged over time, then frequency-binned from 2800 channels to 100, with 2 edge bins removed on each side to exclude any band-edge effects. Figure~\ref{fig: burns law main} shows that the linear polarisation fraction decreases with frequency. The linear polarisation fraction is modelled using Burn's law~\citep{burn1966},
\begin{equation}\label{eq: burn}
    P(\lambda) = e^{-2\sigma_{\rm RM}^2 \lambda^4},
\end{equation}
where \sigmaRM\ quantifies the dispersion in rotation measure along the LOS arising from multipath propagation through a magneto-ionic medium. All fits are performed using weighted least-squares in $\lambda^2$ space. The Burn's-law fit yields \sigmaRM\,$ = 3.56 \pm 0.03$\radm\ (Figure~\ref{fig: burns law main}). 

\section{Discussion}\label{sec: discussion}

\subsection{Implications of component-dependent DM}\label{subsec: DM implications}
While the measured \DM\ values of our FRB sample are consistent with their original \DM{s} within their uncertainties (Section~\ref{subsec: shrine dm}), they clearly look differentially dispersed and the $1\sigma$ allowed \DM\ error range, by eye, looks overly conservative (Figure~\ref{fig: DM error range}). We consider the implications if the apparent offsets between components are genuine. From these offsets, $\Delta\mathrm{DM}_{2-1}\,=\,\DM_2-\DM_1$, we infer differences in the electron column density along their propagation paths using $\Delta n_e \approx \Delta {\rm DM}/d$, where $d \approx c\Delta t$ is the estimated propagation path length and $\Delta t$ is the peak-to-peak component separation measured from the Stokes $I$ pulse profile. This yields density contrasts of order $|\Delta n_e| \sim 10^{9}$--$10^{10}~{\rm cm^{-3}}$ for all three FRBs (Table~\ref{tab:dm_results}). The asymmetric \DM\ errors reported in Table~\ref{tab:dm_results} are propagated to errors in $|\Delta n_e|$ following \citet{2003physics...6138B} and \citet{2019A&A...627A..84L}. Such values require that the two components either probe different regions of the local environment or encounter a medium that evolves on millisecond timescales between their arrival times.

This is naturally explained if the emission regions are spatially separated, so that the paths differ by a characteristic distance $d=c\Delta t$ ($d=\mathcal{O}(100\,{\rm km})$ for the FRBs considered here), allowing each component to sample a different plasma column. Density contrasts of $|\Delta n_e| \sim 10^{9}$--$10^{10}~{\rm cm^{-3}}$ are plausible within neutron star magnetospheres, where the Goldreich-Julian density can reach $\rho_{\rm GJ}\sim10^{12}~{\rm cm^{-3}}$ \citep{goldreich1969}. This picture does not necessarily imply that comparable \DM\ variations should be common in radio pulsars, as their emission is generally believed to arise on open magnetic field lines that largely avoid the dense inner magnetosphere. Similar density variations may arise if one component traverses a dense stellar wind \citep[e.g.,][]{varshni1978}. Wolf-Rayet winds are expected to be highly structured, with 3D simulations finding characteristic scales of ${\sim}0.02\text{-}0.1R_c$, corresponding to $\mathcal{O}(10^4)$--$\mathcal{O}(10^5)\,{\rm km}$ \citep[for $R_c = R_\odot$,][]{moens2022}. Although these structures are larger than the $d=\mathcal{O}(100\,{\rm km})$ path separations inferred here, they demonstrate that substantial density structure can exist on scales far smaller than the overall extent of the wind. By contrast, if the excess dispersion originates in a more distant nebular environment, the inferred densities become difficult to reconcile with known systems; for example, the densest filamentary structures in the Crab Nebula have characteristic electron densities of only $n_e\sim10^{4}~{\rm cm^{-3}}$ \citep{fesen1982,arias2025}. 

Plasma lensing provides another explanation for apparent component-dependent \DM{s}. In addition to the standard cold-plasma dispersive delay, plasma lenses introduce a geometric propagation delay, analogous to gravitational lensing, but with both contributions exhibiting strong frequency dependence \citep[e.g.,][]{cordes2017a}. If the burst is dedispersed assuming only the $\nu^{-2}$ cold-plasma delay, part of the geometric delay ($\propto\nu^{-4}$) is absorbed into the inferred \DM, biasing its measured value \citep{er2020}. \citet{wang2025d} argue that different frequencies traverse different ray paths through the lens, sampling different electron columns and naturally producing frequency-dependent \DM{s}. Recent observations of the repeater FRB~20240619D similarly revealed component-dependent \DM{s} at the level of $\sim0.7\pccm$, with plasma lensing favoured as the preferred explanation \citep{2026MNRAS.546ag090O}.

These effects offer testable observational signatures. Because the geometric delay scales more steeply with frequency than the dispersive delay, broadband or low-frequency observations are best placed to isolate the $\nu^{-4}$ contribution and distinguish it from a genuine \DM\ change \citep{er2020}. The frequency-dependent magnification predicted by \citet{wang2025d} further implies that, for a diverging plasma-lens, the higher-\DM\ component of a lensed pair should coincide with a lower central frequency and lower flux than its counterpart. 

Alternatively, both components may share the same LOS and originate from a common emission region, with \DDM\ arising from temporal evolution of the intervening plasma. The effective electron column between successive components could change through the motion of dense plasma structures across the LOS or through plasma evolution triggered by the first component itself, essentially leading to the same $d=c\Delta t$ constraints described above. In pulsar current sheets, for example, rapidly evolving plasmoids with densities exceeding $\rho_{\rm GJ}$ are expected \citep{philippov2019}. Likewise, rotation of an inhomogeneous magnetosphere or motion of dense wind material could alter the electron column sampled by successive components. Depending on whether an overdense region enters or leaves the LOS between components, either positive or negative \DDM\ could naturally arise. 

If the burst itself modifies the surrounding plasma, the observed \DDM\ may be accompanied by changes in its polarisation properties. \citet{dial2026} show that a sufficiently luminous burst can accelerate free electrons to relativistic speeds, which reduces the effective plasma frequency and suppresses dispersion, Faraday rotation, and scattering for later-arriving emission. In addition, an increased relativistic electron fraction causes the natural wave modes to become increasingly elliptical \citep{kennett1998}, enhancing the efficiency of generalised Faraday rotation (GFR) and linear--circular polarisation conversion. In this picture, differences in \DM\ and polarisation are physically linked, and the presence of circular polarisation in the second component but not the first would provide a potential diagnostic of local plasma evolution. 

\begin{figure}[ht!]
    \centering
    \includegraphics[width=1\linewidth]{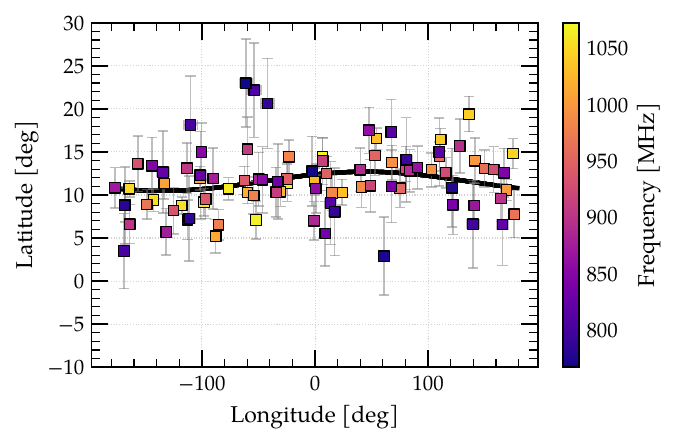}
    \caption{Poincar\'e sphere representation of the polarisation state across $C_2$ of \frb\ prior to \RM\ correction, shown in a rectangular (linear longitude--latitude) projection: longitude is the angle in the $Q$--$U$ plane and latitude the angle out of it. The best-fit small circle (black line) is obtained by variance-weighted least squares on the unit $\{Q,U,V\}$ points (weights $\propto 1/(\sigma_Q^2+\sigma_U^2+\sigma_V^2)$); the fit has residual ${\rm rms} = 0.34^\circ$.}
    \label{fig: poincare freq}
\end{figure}

Qualitatively, \frb\ is consistent with this picture: the second, circularly polarised component has a lower inferred \DM\ than the first (Table~\ref{tab:dm_results}), matching the direction expected if first component drove local electrons relativistic ahead of the second. \frba, however, does not show the same sign of \DDM\ despite also showing circular polarisation preferentially in its second component, disfavouring this mechanism as a universal explanation. Furthermore, neither \frba\ nor \frb\ shows evidence for the characteristic signatures of GFR (Figure~\ref{fig: poincare freq}). While this disfavours strong GFR as the dominant origin of the observed polarisation behaviour, weaker magnetospheric propagation effects that do not produce detectable Poincar\'e signatures cannot yet be excluded (Section~\ref{sec:Component-dependent circular polarisation}), and \DM, faraday rotation and scattering discussed by \citet{dial2026} (see above) remains a further untested possibility given the ambiguity in our component \DM\ offsets (Section~\ref{subsec: shrine dm}).

\subsection{Component-dependent circular polarisation}\label{sec:Component-dependent circular polarisation}

There appears to be an emerging trend among multi-component bursts in the CRAFT sample that circular polarisation is preferentially associated with the second component. Examples include \frba\ (Figure \ref{fig:htr_plot1}), FRB 20211127I, FRB 20230708A, FRB 20240210A \citep[][]{2025arXiv250517497S,dial2025}, and \frb\ (Figure \ref{fig: RM}). The trend is not universal, however, with some bursts showing circular polarisation in the first component \citep[e.g., FRB 20240201][]{2025arXiv250517497S}. Furthermore, repeaters are known to exhibit substantial burst-to-burst variations in their polarimetric properties \citep[e.g.,][]{2022SciBu..67.2398F,2023PhRvD.108d3009K,jiang2025}, and it remains unclear whether the apparent preference for circular polarisation in the second component represents a genuine population characteristic or simply reflects the small number of available bursts.

The origin of circular polarisation in FRBs is also uncertain. Circular polarisation may be intrinsic to the emission process or generated through propagation effects either within the magnetosphere of a neutron star or in the surrounding environment \citep[e.g.,][]{kennett1998,mckinnon2024}. Nevertheless, the component-dependent behaviour provides some clues. If the observed circular polarisation is intrinsic, then the presence of strong Stokes $V$ in one component but not the other suggests that the two components originate from physically distinct regions or are produced by different emission conditions. Conversely, if circular polarisation is produced through propagation effects, then the two components must encounter different plasma conditions along their propagation paths. The component-dependent behaviour provides similar constraints to those inferred from the observed \DM\ offsets (Section~\ref{subsec: DM implications}). If the circular polarisation is intrinsic, then differences in Stokes $V$ imply different emission conditions between components. Alternatively, if circular polarisation is generated through propagation, then the components must encounter different plasma conditions along their respective paths.

Neither \frba\ nor \frb\ show clear evidence for GFR. In both bursts, the Poincar\'e frequency spectra of the circularly polarised components (e.g., Figure~\ref{fig: poincare freq}) do not exhibit the characteristic great- or small-circle trajectories expected for GFR \citep[e.g.,][]{kennett1998}. Following \citet{noutsos2009}, we searched for signatures of magnetospheric propagation expected if the observed polarisation evolution is produced by GFR. In this scenario, the greatest $\Delta\Pi_V(\nu)$ --- the change in $\Pi_V$ with frequency --- is expected to coincide with the greatest deviation of the apparent time-resolved \RM\ from its mean value \citep[][Section~\ref{subsec: RM}]{noutsos2009}. However, \citet{ilie2019} argue that scattering of a burst with an intrinsic PA swing can produce qualitatively similar behaviour, complicating attempts to distinguish between these mechanisms observationally. We did not detect any significant correlation owing to the limited Stokes $V$ S/N. Figures~\ref{fig: burns law main} and \ref{fig: poincare freq} show that $\Pi_V$ is roughly constant with frequency, however, this may be expected for pair-plasma magnetar winds causing birefringence/GFR close to the source~\citep[][]{lyutikov2022}.

The absence of these signatures does not exclude propagation effects entirely, and the component-dependent Stokes $V$ may reflect intrinsic differences between the two emitting regions or propagation through plasma structures that affect only one component. Several mechanisms capable of producing circular polarisation therefore remain viable. Intrinsic emission models, such as coherent curvature radiation or inverse Compton scattering, can produce component-dependent Stokes $V$ if the two components originate from regions with different magnetic field orientations or viewing geometries \citep[][]{qu2023b}. Alternatively, propagation effects such as cyclotron absorption or plasma mode conversion may introduce frequency-dependent circular polarisation signatures \citep[][]{wang2010,qu2022}. Future broadband polarimetric observations of repeating multi-component bursts will be crucial for distinguishing between these scenarios by tracking the frequency evolution of Stokes $V$, its relationship with burst intensity and spectral structure, and its variation between components. Such observations will determine whether the apparent preference for circular polarisation in the second component reflects differences in the emission process or in the intervening plasma. Regardless of its origin, the observed polarisation is further modified by propagation through magneto-ionic media. We therefore next consider time-resolved RM variations and the role of scattering in shaping the observed PA evolution.

If the component-dependent Stokes~$V$ is intrinsic, a natural candidate is that successive components sample different viewing-angles through an inhomogeneous or evolving magnetosphere, since the degree of circular polarisation produced by curvature radiation or inverse Compton scattering is sensitive to the line-of-sight angle relative to the local magnetic field \citep[][]{qu2023b}. Such a geometric change between components would be expected to leave a signature not only in $\Pi_V$ but also in the polarisation angle and, if the components probe distinct plasma columns, in \DM. This picture offers a self-consistent link between three otherwise independent observations in our sample: the component-dependent Stokes~$V$ seen in \frb\ and \frba, the $\sim20^\circ$ PA jump between components of \frbb\ (Section~\ref{subsec: time RM PA}), and the apparent \DDM\ offsets discussed in Section~\ref{subsec: DM implications}.

\subsection{PA morphology and time-resolved RM}\label{subsec: time RM PA}


The persistence of the PA swing through the scattering tail of \frb\ is unusual (Figure~\ref{fig: RM}). Scattering is expected to flatten an intrinsic PA swing because the observed emission at a given time is a superposition of radiation arriving along paths with different delays and therefore different intrinsic PAs. This behaviour has been demonstrated for pulsars \citep[e.g.,][]{2003A&A...410..253L,karastergiou2009}, and is expected to apply to FRBs as well \citep[e.g.,][]{caleb2018,balzan2026}. \citet{2025arXiv250517497S} show this; FRBs with greater scattering timescales show less structure in their PAs (see their Figure 10). The persistence of a coherent PA swing through the scattering tail of \frb\ is therefore difficult to reconcile with scattering. 

\begin{figure}[ht!]
    \centering
    \includegraphics[width=1\linewidth]{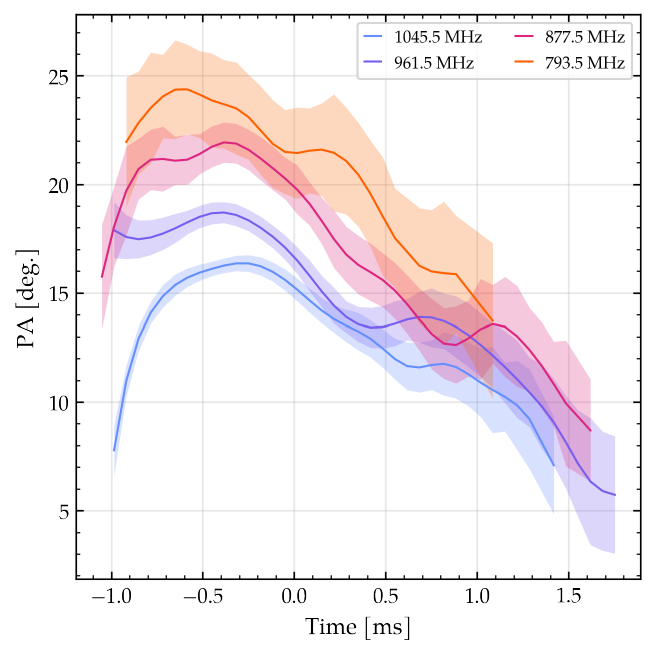}
    \caption{\frb\ $C_2$ (decimated in time by a factor of four) PA profiles in four equal-bandwidth subbands spanning the observing band.}
    \label{fig: PA freq}
\end{figure}

This discrepancy is also evident in the frequency dependence of the PA morphology. Because scattering is stronger at lower frequencies, scattering of an intrinsic PA swing should produce greater PA flattening at lower frequencies. Instead, the PA swing of \frb\ seems to become progressively flatter towards higher frequencies (Figure~\ref{fig: PA freq}), opposite to the behaviour expected from scattering alone. We are not aware of a comparable observation of substantial pulse broadening accompanied by a coherently preserved PA structure with this frequency dependence.

One possible interpretation is that scattering occurs while the PA is still evolving within the near-source magnetospheric environment. In this picture, the radiation is scattered while it is still propagating through a region in which its polarisation state is changing. The resulting scattered rays could therefore retain information about the evolving PA rather than simply averaging over a completed intrinsic PA swing. Such a configuration has, to our knowledge, not previously been considered in the context of FRB scattering, and it is unclear whether the required physical conditions can be realised in a neutron-star magnetosphere. In particular, the scattering material would need to produce the observed millisecond-scale delays while remaining coupled to the region responsible for the PA evolution. We therefore regard this interpretation as speculative. The absence of similar observations may indicate that such a configuration requires an unusually favourable geometry or extreme plasma conditions. 

An alternative is that the scattering and PA evolution occur in physically distinct environments but in an unusual order along the LOS. For example, the burst could first be scattered in the host galaxy and subsequently pass through the magnetosphere of a rotating neutron star. If the ray path intersected an open magnetic field region, perhaps in a binary system~\citep[e.g., a Be star,][]{dial2026}, magnetospheric propagation could then rotate the PA after the scattering had occurred, naturally allowing a PA swing to be superimposed on a scattered pulse. Although the probability of such an alignment is expected to be very small, this scenario illustrates that a PA swing observed in a scattered pulse does not necessarily require the scattering and polarisation evolution to occur in the same physical region. This scenario is, in principle, testable: if the PA swing arises from traversal of an ordered dipolar field near a magnetic pole, it may be described by the rotating vector model \citep{radhakrishnan1969}.

A further possibility is that the scattering medium itself possesses an ordered magnetic field. A scattering medium with a strongly disordered magnetic field could introduce different Faraday rotations along different propagation paths, tending to depolarise the scattered emission if these path-to-path \RM\ variations are sufficiently large (Equation~\eqref{eq: burn}). Conversely, an ordered magnetic field could allow systematic differences in Faraday rotation between the paths. Because these paths also have different geometric delays, the relative contribution of different Faraday-rotated paths changes across the scattering tail, potentially producing an effective time-dependent \RM. This interpretation also requires care in distinguishing the physical \RM\ from the \RM\ inferred from the observed PA. As discussed in \ref{app:RM scatter}, Faraday rotation from multiple screens remains additive at the level of the polarisation signal, but scattering can introduce an additional apparent \RM\ through the frequency-dependent mixing of emission with different intrinsic PAs. A time-resolved \RM\ therefore need not correspond directly to a changing physical \RM\ along the line of sight on the order of milliseconds. Nevertheless, if the observed time-dependent \RM\ is primarily produced by Faraday rotation associated with the scattering paths, derotating each time segment by its measured \RM\ should remove the corresponding frequency-dependent PA rotation and reveal the PA evolution expected from scattering alone.

\begin{figure}[ht]
    \centering
    \includegraphics[width=\linewidth]{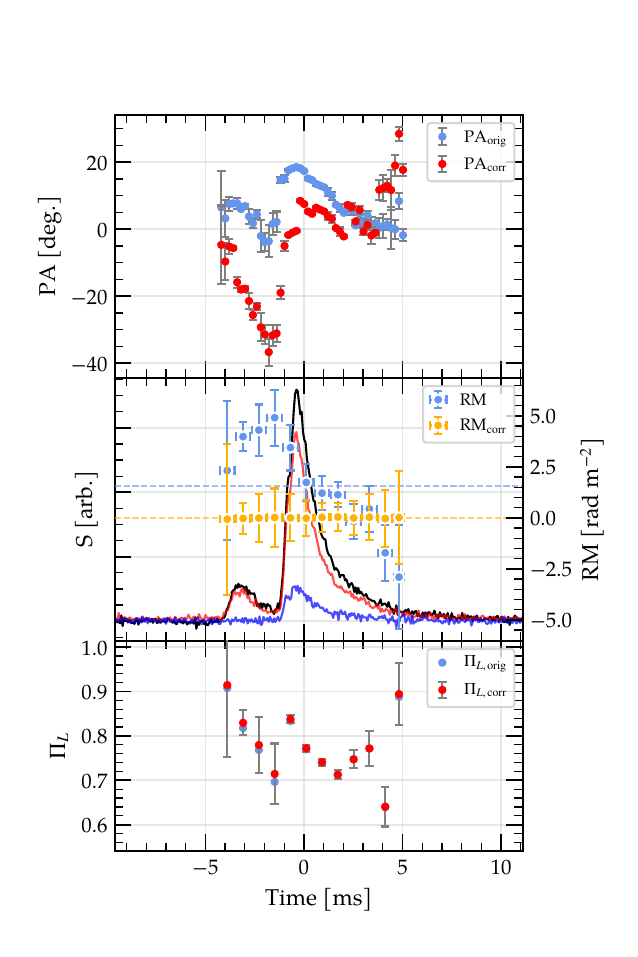}
    \caption{As Figure~\ref{fig: RM}, after reconstructing the dynamic spectrum using the best-fitting time-resolved \RM\ values.}
    \label{fig: dRM derot}
\end{figure}

We tested this by reconstructing \frb\ after derotating 12 time segments by its individually measured \RM\ (Figure~\ref{fig: dRM derot}). If the observed PA evolution were dominated by a time-dependent Faraday rotation introduced along the propagation paths, this correction should remove the corresponding frequency-dependent PA rotation, and the corrected profile should approach the flattened PA morphology expected. Instead, the correction introduces a pronounced PA excursion of $\sim+30^\circ$ towards the end of the scattering tail. $C_1$ likewise does not recover the flattened morphology expected from scattering alone. The persistence of substantial PA structure after correction shows that a simple effective-\RM\ description of the scattering screen is insufficient; a magnetised screen is not excluded, but testing it properly requires modelling scattering and Faraday rotation jointly rather than per-segment.

Taken together, none of the candidate mechanisms considered above provide a complete account of the observed behaviour of \frb\ on its own. The persistence of the PA swing through the scattering tail argues against a simple scattering model, while the derotation experiment indicates that a purely time-dependent Faraday screen cannot fully reproduce the observed PA morphology. The magnetosphere scenario remains geometrically improbable but cannot be excluded, and the near-source coupling scenario remains speculative pending a viable physical mechanism. Conversely, we find no convincing evidence for the characteristic polarisation signatures expected from GFR (Section~\ref{sec:Component-dependent circular polarisation}; Figure~\ref{fig: poincare freq}), although the available Stokes~$V$ sensitivity limits the strength of this conclusion. We therefore favour an interpretation in which the observed time-resolved \RM\ reflects a combination of scattering, intrinsic emission geometry, and possibly magnetospheric propagation effects, rather than any single process acting in isolation.

\subsection{A two-screen scattering geometry}\label{sec:two screen}
The scattering and scintillation properties of \frb\ suggest that the observed scattering and scintillation bandwidth originate from distinct propagation screens (Section~\ref{sec:Scintillation Bandwidth}, \ref{app:NE2025}). We therefore consider a two-screen scattering geometry following \citet{pradeepe.t.2025a}, where the observed scattering is produced by a host-galaxy screen while the scintillation bandwidth is dominated by a Galactic screen. The two-screen formalism relates the source-side and observer-side screen distances through
\begin{equation}\label{eq:Pradeep2ScreenScat}
    D_{h,\rm FRB}\,D_{\rm MW} \lesssim \frac{(1+z)\,D_{\rm FRB}^2}{8\pi\nu^2}\,\frac{\Delta\nu_{\rm d}}{m_g\,\tau},
\end{equation}
where $D_{h,\rm FRB}$ is the distance between the FRB source and the host scattering screen, $D_{\rm MW}$ is the distance between the Milky Way scattering screen and the observer, and $D_{\rm FRB}$ is the angular diameter distance to the source. Using the measured values of $\Delta\nu_{\rm d}=0.7113$\,MHz, $\tau=1.25$\,ms, and $m_g=0.3014$, we obtain
\begin{equation}
    D_{h,\rm FRB}\,D_{\rm MW} \lesssim 57.6~{\rm kpc}^2 .
\end{equation}
Adopting the effective Galactic screen distance inferred from NE2025, $D_{\rm MW}=D_{\rm MW,eff}\simeq0.9392$\,kpc (Section~\ref{sec:Scintillation Bandwidth}), gives the upper limit on the distance between the FRB source and the host scattering screen:
\begin{equation}
    D_{h,\rm FRB} \lesssim \frac{57.6~{\rm kpc}^2}{0.9392~{\rm kpc}} \simeq 61.3~{\rm kpc},
\end{equation}
placing the dominant scattering screen within the host-galaxy.\footnote{Prior formulations of this constraint by \citet{2022ApJ...931...87O} and \citet{2023MNRAS.525.5653S} differ from Equation~\eqref{eq:Pradeep2ScreenScat} in two respects, as identified by \citet{pradeepe.t.2025a}: the redshift factor enters the numerator rather than the denominator (see their Section~2, footnote~1), and the modulation index enters linearly following Equations~7.4--7.5 of \citet{pradeepe.t.2025a}, rather than as $m_g^{-2}$. The $m_g^{-2}$ scaling of \citet{2023MNRAS.525.5653S} yields a substantially weaker upper limit, $D_{h,\rm FRB}\lesssim555$\,kpc, for \frb.} Taken together, the derived limit of $D_{h,\rm FRB}\lesssim61$\,kpc constrains the dominant scattering screen to lie within the host galaxy's interstellar medium, halo, or immediate circumburst environment.

\subsubsection{Origin of the low modulation index.}\label{sec: low m}
We do not attribute the low modulation index to weak scintillation, as the measured scintillation bandwidth satisfies $\Delta\nu_d/\nu \ll 1$, indicating that the observations are firmly in the strong-scattering regime. The observed suppression of $m_g$ below unity may be understood in terms of the resolution power (RP) formalism of \citet{pradeepe.t.2025a}. When two scattering screens are separated by a distance $D_{\rm MW,host}$, each screen's ability to resolve the other is set by the ratio of the second screen's angular size to the first screen's diffraction-limited angular resolution. \citet{pradeepe.t.2025a} define this ratio as
\begin{equation}\label{eq:RP-def}
    \mathrm{RP} = \frac{L_{\rm MW}L_{\rm host}}{\lambda D_{\rm MW,host}},
\end{equation}
where $L_{\rm MW}$ and $L_{\rm host}$ are the physical transverse sizes (effective apertures) of the Milky Way and host-galaxy screens, respectively, and $\lambda$ is the observing wavelength. Screens with $\mathrm{RP} \ll 1$ do not resolve each other and produce the single-screen modulation index $m_g=1$; as $\mathrm{RP}$ increases past unity, each screen increasingly resolves the other, progressively suppressing and broadening the corresponding scintillation component. This suppression follows \citep{pradeepe.t.2025a}
\begin{equation}\label{eq:mg-RP-invert}
    m_g = \left(1+\frac{\pi^2}{4^3}\mathrm{RP}^2\right)^{-1/2}
    \;\;\Rightarrow\;\;
    \mathrm{RP} = \frac{8}{\pi}\sqrt{\frac{1}{m_g^2}-1}.
\end{equation}
For our measured $m_g=0.3014$, Equation~\eqref{eq:mg-RP-invert} gives $\mathrm{RP}\approx8.06$, placing the system well within the resolving regime.
As a check on the physical plausibility of this scenario, we use $\mathrm{RP}$ to estimate the implied sizes of both screens. Because $\mathrm{RP}\gg1$, the observed $\Delta\nu_{d,\rm MW}$ is itself broadened relative to the intrinsic Milky Way screen size \citep{pradeepe.t.2025a}:
\begin{align}\label{eq:nu-broadened}
    \Delta\nu_{d,\rm MW} &= \frac{c}{\pi D_{\rm MW}\theta_{L,\rm MW}^2}
    \sqrt{1+\frac{\pi^2}{4^3}\mathrm{RP}^2}\\
    &= \frac{c}{\pi D_{\rm MW}\theta_{L,\rm MW}^2}\,\frac{1}{m_g}.
\end{align}
Solving for $\theta_{L,\rm MW}$ and substituting into the screen-size definition $L_{\rm MW}=4\theta_{L,\rm MW}D_{\rm MW}$ \citep{pradeepe.t.2025a} gives
\begin{equation}\label{eq:LMW}
    L_{\rm MW} = 4\sqrt{\frac{c\,D_{\rm MW}}{\pi\,\Delta\nu_{d,\rm MW}\,m_g}}.
\end{equation}
Adopting the NE2025-inferred Galactic screen distance, $D_{\rm MW}\simeq0.9392$\,kpc (\ref{app:NE2025}), Equation~\eqref{eq:LMW} gives
\begin{equation}
    L_{\rm MW} \approx 3.04~\mathrm{AU},
\end{equation}
consistent with AU-scale Galactic scattering disks \citep[e.g.,][]{brisken2010,kerr2018}. As an independent check, applying the extragalactic angular-broadening relation of \citet[][their Eq.~8]{cordes2002} directly to the NE2025-predicted extragalactic angular-broadening scattering measure ($\mathrm{SM}_{\theta,x}$; Appendix~\ref{app:host screen}) gives $L_{\rm MW}\approx3.43$\,AU, in good agreement with the value derived from the measured $m_g$ and $\Delta\nu_{d,\rm MW}$. We adopt $L_{\rm MW}\approx3.04$\,AU as our fiducial value. The corresponding host screen size then follows directly from 
Equation~\eqref{eq:RP-def}:
\begin{equation}
    L_{\rm host} = \frac{\mathrm{RP}\,\lambda\,D_{\rm MW,host}}{L_{\rm MW}},
\end{equation}
at the burst reference frequency ($\nu_{\rm obs}=919.5$\,MHz). We adopt $D_{\rm MW,host}\simeq D_{\rm FRB} = 7.318\times10^5$\,kpc. Substituting $\mathrm{RP}\approx8.06$ and $L_{\rm MW}\approx3.04$\,AU gives
\begin{equation}
    L_{\rm host} \approx 8.7\times10^{2}~\mathrm{AU}.
\end{equation}
This is one to two orders of magnitude larger than the transverse scales of well-characterised Galactic scattering structures. For comparison, VLBI observations of PSR B0834+06 revealed a scattering structure $\sim16$ AU in extent \citep{brisken2010}. While screen sizes in host-galaxy or circumgalactic environments need not match Galactic ISM values, we regard it as a caveat to the resolving interpretation rather than a confirmation of it.

\citet{pradeepe.t.2025a} also show that an equivalent broadening and suppression of $m_g$ arises if a single screen resolves an intrinsically incoherent emission region, rather than a second scattering screen \citep[see also][]{kumar2024}; the two scenarios can, in principle, be distinguished through the frequency dependence of the scintillation bandwidth, which we are unable to constrain reliably in this burst owing to limited bandwidth and spectral S/N (Section~\ref{sec:Scintillation Bandwidth}). This limitation is common across the CRAFT FRB sample \citep{2023MNRAS.525.5653S}. Equation~\eqref{eq:Pradeep2ScreenScat} therefore constrains the host-screen distance under an assumed two-screen geometry, rather than stating definitive evidence for a second, independently detected scattering screen.

We consider three possible instrumental or intrinsic origins for the low $m_g$ and argue that neither can account for the observed suppression. First, self-noise has already been excluded as a contributor to the measured spectral feature (Section~\ref{sec:Scintillation Bandwidth}); in any case, self-noise from an unresolved intrinsic pulse shape introduces an additional, effectively multiplicative contribution to the spectral ACF that inflates the peak correlation above the single-screen value, rather than suppressing it \citep{pradeepe.t.2025a}, so its presence could not account for $m_g<1$ even if it were relevant here. Second, dilution from intra-pulse evolution of the scintillation pattern --- expected in a resolving two-screen system, where $m_g$ is predicted to decrease and $\Delta\nu_{\rm d}$ to increase across the pulse profile \citep{pradeepe.t.2025a} --- can also suppress a burst-averaged $m_g$ below its instantaneous value. However, the burst-averaged modulation index measured from the FWHM-spectrum ACF (Section~\ref{sec:Scintillation Bandwidth}), $m_g=0.3014$, is consistent with the peak of the time-resolved modulation index, $m_g=0.33$ (Figure~\ref{fig: mod idx}), indicating that averaging over the pulse profile does not substantially bias our measurement. The low modulation index therefore does not appear to be an artefact of noise or pulse-averaging. Further, as shown by L. Nicotera et al. (in prep, 2026), calculated modulation indices may differ from the expected $m_g=1$, when small numbers of scintils are observed. However, when the number of scintils is large (here, $336/0.71 \sim 473$), the calculated value of $m_g$ tends to vary by $\sim10\%$ --- thus, we exclude that the true value of $m_g$ is 1. 

\subsubsection{\sigmaRM-$\tau$ relation}
\begin{figure}[ht!]
    \centering
    \includegraphics[width=1\linewidth]{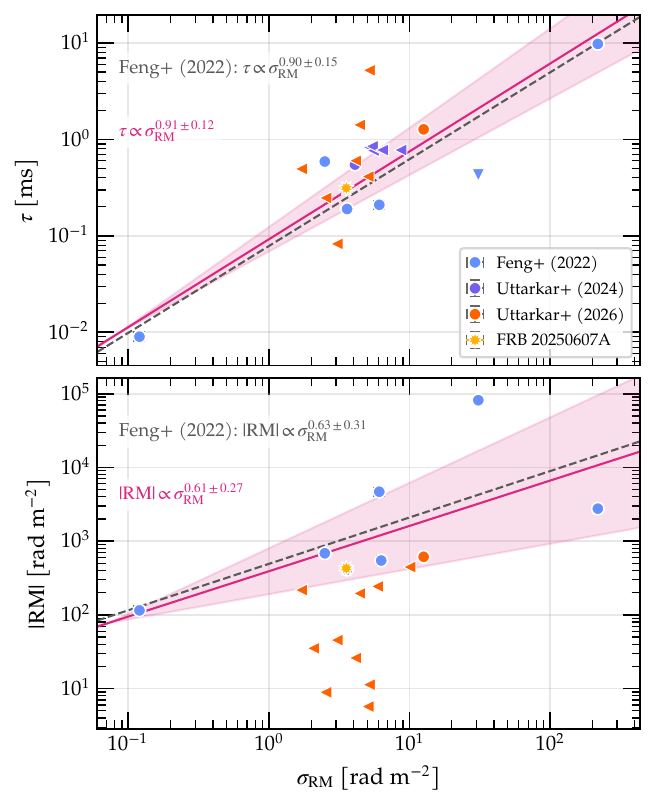}
    \caption{$\sigma_{\rm RM}$ versus $\tau$ and $|\RM|$. Blue points are repeating FRBs used in \citet{feng2022b} (see references therein), purple and orange points are apparent non-repeating CRAFT FRBs from \citet{uttarkar2024a} and \citet{uttarkar2026}, respectively, and the yellow star marks \frb. Top: scattering timescale as a function of $\sigma_{\rm RM}$. Power-law fits of the form $\tau_{\rm 1.3,GHz}\propto\sigma_{\rm RM}^{\alpha}$ are shown for the \citet{feng2022b} sample (grey dashed) and the full sample including \frb\ (magenta solid, with the shaded region indicating the $1\sigma$ uncertainty). The fitted slopes are $\alpha=0.90\pm0.15$ for the \citet{feng2022b} sample and $\alpha=0.91\pm0.12$ for the full sample. Bottom: $|\RM|$ as a function of $\sigma_{\rm RM}$, with corresponding power-law fits $|\RM|\propto\sigma_{\rm RM}^{\beta}$. We obtain $\beta=0.63\pm0.31$ for the \citet{feng2022b} sample and $\beta=0.62\pm0.27$ for the full sample. Downward triangles indicate upper limits on $\tau$, while left-pointing triangles indicate upper limits on $\sigma_{\rm RM}$. All scattering times are scaled to 1300 MHz assuming $\tau\propto\nu^{-4}$, except for \frb, for which $\tau\propto\nu^{-4.19}$ is adopted.}
    \label{fig: rm tau}
\end{figure}

As a further check on the physical plausibility of the inferred host-screen environment, the measured \RM\ scatter provides independent evidence for magnetised multipath propagation. We measure $\sigma_{\rm RM}=3.56\pm0.03~\mathrm{rad,m^{-2}}$ (Section~\ref{subsec: RM}), which is consistent with the range associated with multipath propagation through magnetised, turbulent plasma screens. \citet{feng2022b} reported a correlation between $\sigma_{\rm RM}$ and scattering timescale, finding $\tau_{\rm 1.3,GHz}\propto\sigma_{\rm RM}^{0.81\pm0.16}$, while \citet{yang2022a} reproduced a similar scaling theoretically. To compare our measurements with the \citet{feng2022b} relation, we independently fit their sample using ordinary least-squares regression in log space (Figure~\ref{fig: rm tau}). This gives $\tau\propto\sigma_{\rm RM}^{0.90\pm0.15}$, somewhat steeper than their reported exponent, likely reflecting the fitting procedure: \citet{feng2022b} do not specify sufficient details of their fitting method. Applying the same fitting procedure to the combined \citet{feng2022b} sample and the additional CRAFT non-repeating FRBs~\citep[][]{uttarkar2024a,uttarkar2026} gives $\tau\propto\sigma_{\rm RM}^{0.91\pm0.12}$, while the corresponding relation between absolute \RM\ and \RM\ scatter is $|\mathrm{RM}|\propto\sigma_{\rm RM}^{0.63\pm0.31}$ for the \citet{feng2022b} sample and $|\mathrm{RM}|\propto\sigma_{\rm RM}^{0.62\pm0.27}$ for the combined sample. The inclusion of the newer CRAFT measurements does not significantly alter either scaling. These correlations provide circumstantial support for a magnetised scattering environment, although they do not by themselves establish that the inferred host screen is responsible for the time-resolved \RM\ structure observed in \frb.

\section{Conclusions}\label{sec: conclusions}

We have presented a propagation and polarimetric analysis of \frb, a two-component FRB detected by CRAFT, and compared its component-dependent properties with two further CRAFT bursts. Our main results point towards a common origin for several otherwise unusual observables.

Our \DM\ optimisation finds no statistically significant differential \DM\ in FRB 20250607A, FRB 20190611B, or FRB 20210407E, despite apparent component-dependent frequency structure. The formal uncertainties may therefore be conservative, although low-level differential \DM\ cannot be excluded. If the apparent offsets are real, the implied electron-density contrasts of $|\Delta n_e|\sim10^{9}$--$10^{10}\,\pccm$ favour plasma local to the FRB source, such as an inhomogeneous or evolving neutron-star magnetosphere, rather than ordinary interstellar plasma. 

An investigation of the structure-maximising technique and its associated filter design \citep{2023ApJ...954...37S} and comparison with polarimetric \DM\ optimisation metrics will be presented in a future work (Balzan et al., in prep.).

The most striking constraint comes from combining the scattering and polarimetric behaviour of \frb. Its scattering timescale and scintillation bandwidth require at least two propagation screens. Yet the PA swing persists through the scattering tail rather than flattening as expected for simple scattering --- to our knowledge, no comparable case of substantial pulse broadening coexisting with a coherently preserved PA swing has previously been reported. Together with the time-resolved \RM\ variations and component-dependent circular polarisation, this suggests that the observed scattering cannot be treated as an entirely independent, post-emission propagation effect. Instead, the polarisation-changing magnetospheric plasma and the scattering material may be spatially or temporally coupled, raising the possibility that significant scattering occurs while the radiation is still propagating through the structured near-source environment.

The component-dependent \DM, circular polarisation, and PA structure are therefore plausibly manifestations of differing sightlines through an inhomogeneous or evolving magnetosphere. We find no evidence for strong generalised Faraday rotation, so the data do not require a single specific propagation mechanism; rather, they point to a complex near-source environment in which emission, polarisation evolution, and scattering may be interconnected.

Future observations of multi-component FRBs with high time and frequency resolution, broad fractional bandwidth, and sensitive full-Stokes measurements will be critical for testing this picture. In particular, simultaneous measurements of component-dependent \DM, scattering, scintillation, and polarisation across a wide frequency range can determine whether these signatures systematically co-occur and whether scattering is indeed coupled to magnetospheric propagation.

\begin{acknowledgement}
This scientific work uses data obtained from Inyarrimanha Ilgari Bundara, the CSIRO Murchison Radio-astronomy Observatory. We acknowledge the Wajarri Yamaji People as the Traditional Owners and native title holders of the Observatory site. CSIRO's ASKAP radio telescope is part of the Australia Telescope National Facility (https://ror.org/05qajvd42). Operation of ASKAP is funded by the Australian Government with support from the National Collaborative Research Infrastructure Strategy. ASKAP uses the resources of the Pawsey Supercomputing Research Centre. Establishment of ASKAP, Inyarrimanha Ilgari Bundara, the CSIRO Murchison Radio-astronomy Observatory and the Pawsey Supercomputing Research Centre are initiatives of the Australian Government, with support from the Government of Western Australia and the Science and Industry Endowment Fund.\\

This work was performed on the OzSTAR national facility at Swinburne University of Technology. The OzSTAR program receives funding in part from the Astronomy National Collaborative Research Infrastructure Strategy (NCRIS) allocation provided by the Australian Government, and from the Victorian Higher Education State Investment Fund (VHESIF) provided by the Victorian Government.\\

This research was supported by an Australian Government Research Training Program (RTP) Scholarship (\url{https://doi.org/10.82133/C42F-K220}).
AB acknowledges support through project CORTEX (NWA.1160.18.316) of the research programme NWA-ORC which is financed by the Dutch Research Council (NWO). 
ATD and RMS acknowledge support through Australian Research Council Discovery Project DP220102305.
\end{acknowledgement}

\section*{Data Availability}
ASKAP data for \frba\ \frbb\ are available online at \url{https://doi.org/10.25917/1RG2-C612}. The \texttt{FIRES} codebase is available at \url{https://github.com/JoelBalzan/FIRES}. Codes used for \DM\ optimisation and all other analyses are available at \url{https://github.com/JoelBalzan/FRBop}. The data used in this work will be made available upon reasonable request.

\bibliography{bibliography}

\appendix

\section{DM optimisation}
\begin{figure}[ht!]
    \centering
    \begin{subfigure}[t]{\linewidth}
        \centering
        \includegraphics[width=0.9\linewidth]{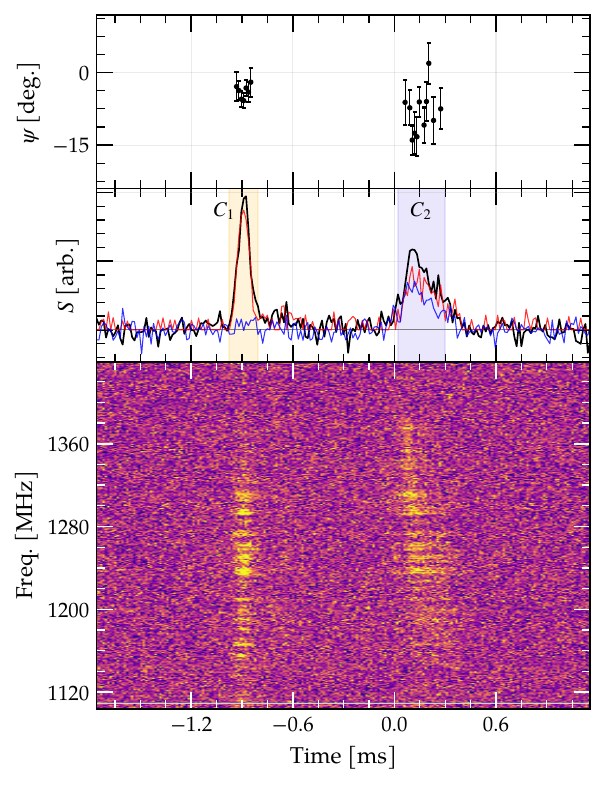}
        \caption{\frba\ with time resolution $\delta t = 0.014018$\,ms and frequency resolution $\delta\nu = 1$\,MHz.}
        \label{fig:htr_plot1}
    \end{subfigure}
    
    \begin{subfigure}[t]{\linewidth}
        \centering
        \includegraphics[width=0.9\linewidth]{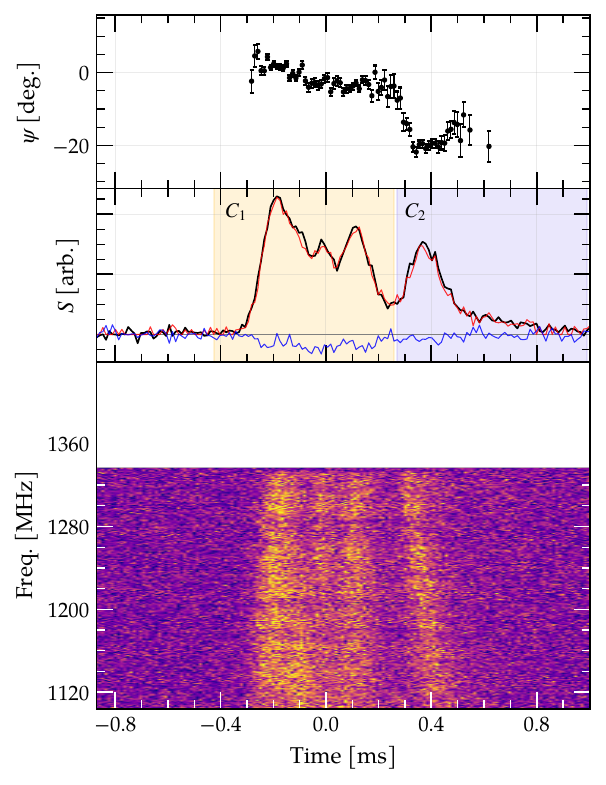}
        \caption{\frbb\ with time resolution $\delta t = 0.012016$\,ms and frequency resolution $\delta\nu = 1$\,MHz.}
        \label{fig:htr_plot2}
    \end{subfigure}
    \caption{
    High-time-resolution plots for \frba\ and \frbb. Panels are described in Figure~\ref{fig:htr_plot3}. 
    }
    \label{fig:htr_plots}
\end{figure}

\begin{figure*}[ht]
    \centering
    \includegraphics[width=1\linewidth]{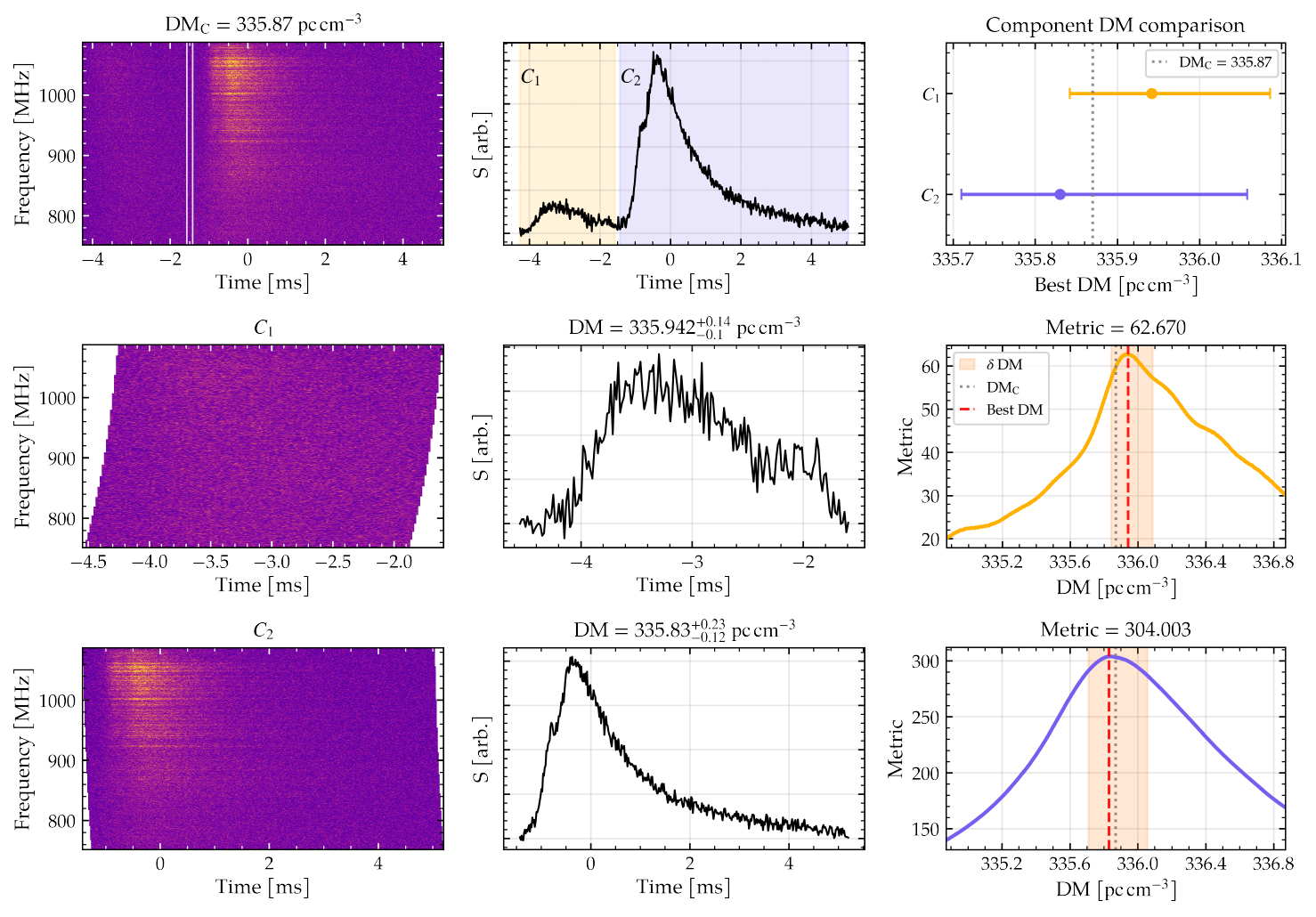}
    \caption{Comparison of the structure-maximising (\DM$_{\rm struct}$) optimisation for the two components of \frb\ over the range \DM\,=\,334.87--336.87~\pccm. The top row shows the original burst dedispersed at the CELEBI value, \DM$_{\rm C}=335.87$~\pccm, with white vertical lines and shaded regions corresponding to the component windows. The second and third rows show the results for the first and second burst components, respectively. The left column shows the incoherently dedispersed dynamic spectrum at the best-fit \DM, the middle column shows the corresponding frequency-summed Stokes $I$, and the right column shows the structure-maximisation metric profile. In each scan, the CELEBI \DM\ is indicated by the dotted line and the best-fit \DM$_{\rm struct}$ by the red dashed line.}
    \label{fig:250607_dm_structure}
\end{figure*}

\begin{figure*}[ht!]
    \centering
    \includegraphics[width=\linewidth]{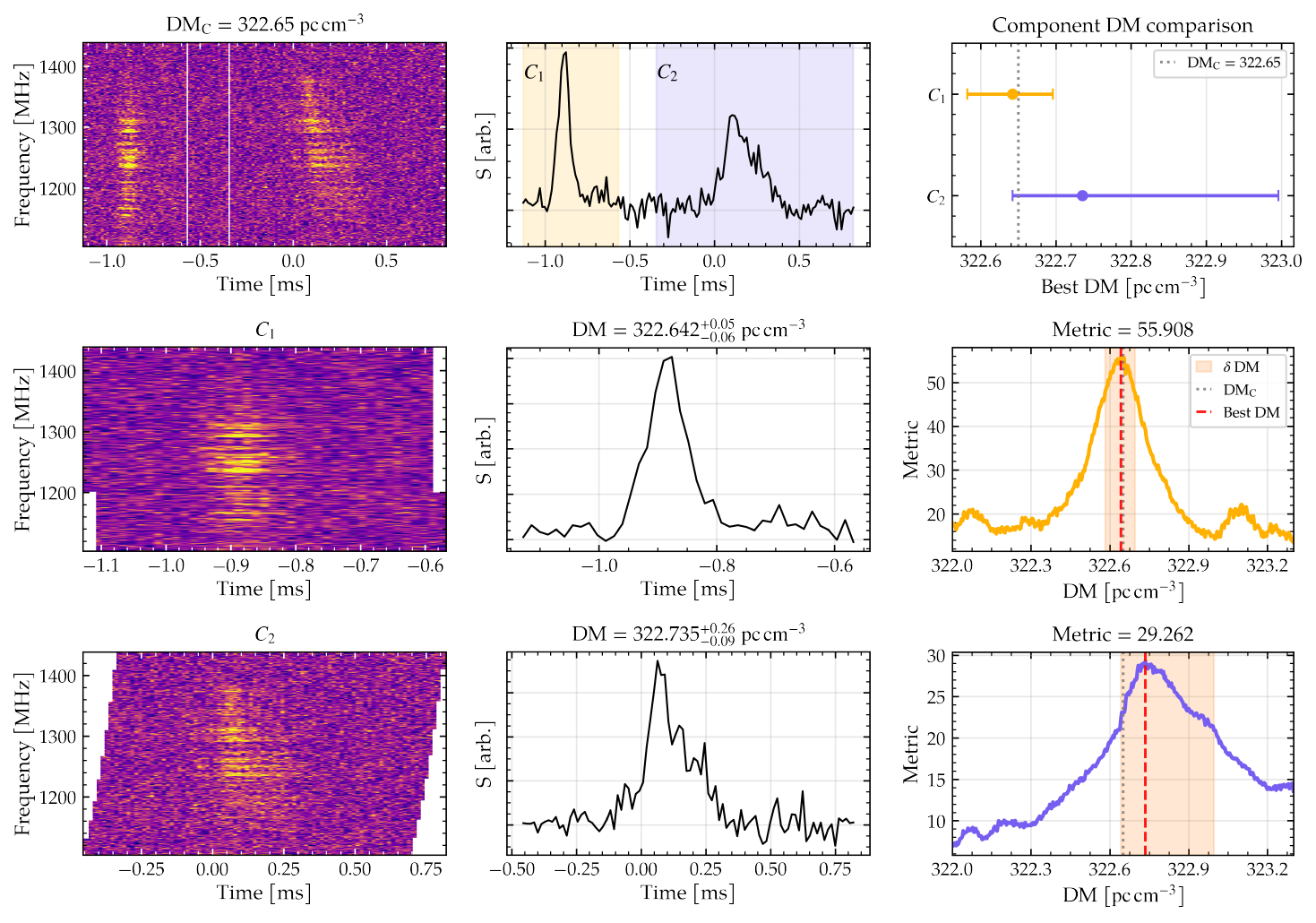}
    \caption{Comparison of the structure-maximising (\DM$_{\rm struct}$) optimisation for the two components of \frba. The top row shows the original burst dedispersed at the CELEBI value, \DM$_{\rm C}=322.65$~\pccm. Panels are described in Figure~\ref{fig:250607_dm_structure}.}
    \label{fig:190611_dm_structure}
\end{figure*}

\begin{figure*}[ht!]
    \centering
    \includegraphics[width=\linewidth]{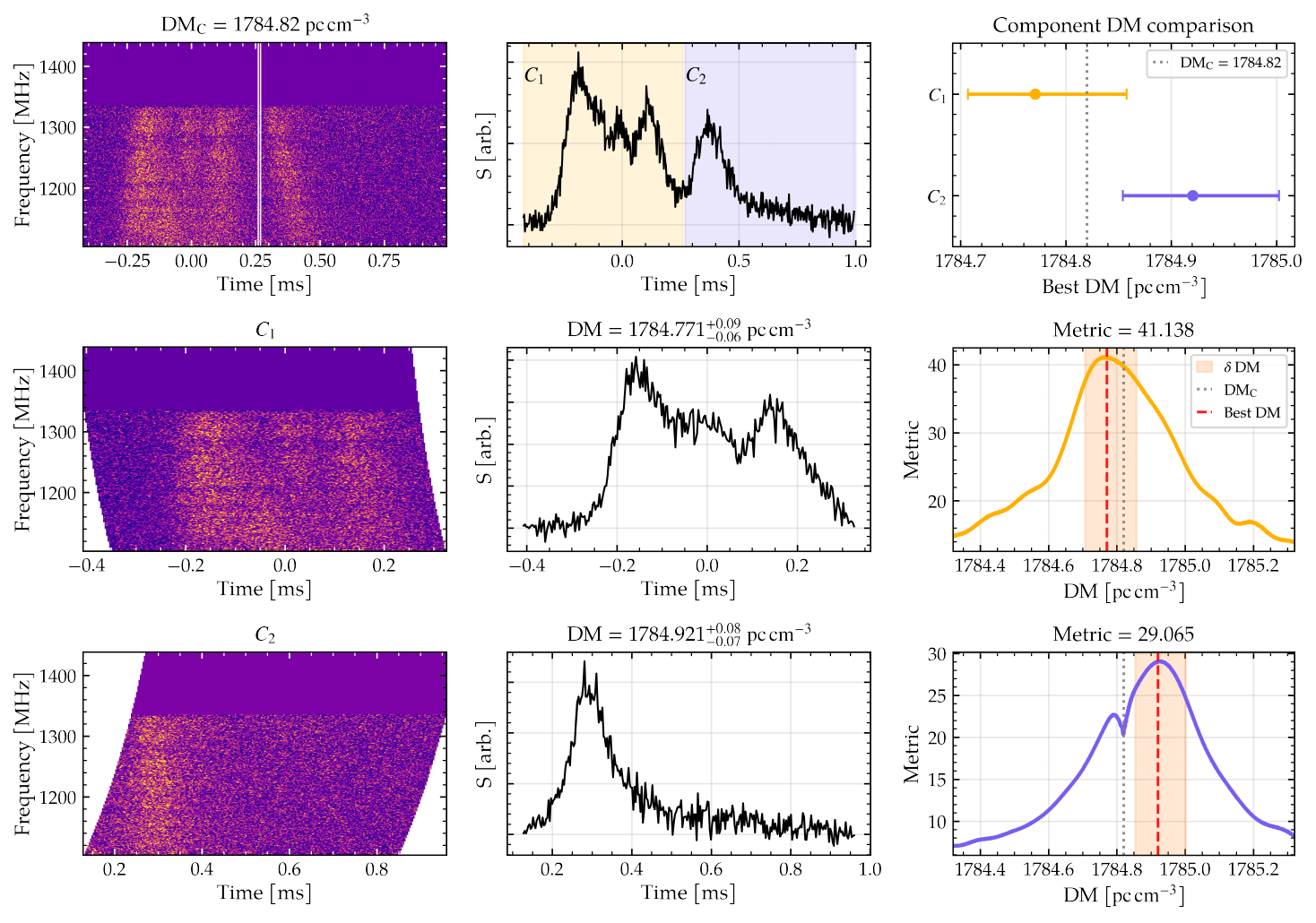}
    \caption{Comparison of the structure-maximising (\DM$_{\rm struct}$) optimisation for the two components of \frbb\ at time resolution $\delta t = 0.003$\,ms. The top row shows the original burst dedispersed at the CELEBI value, \DM$_{\rm C}=1784.82$~\pccm. Panels are described in Figure~\ref{fig:250607_dm_structure}.}
    \label{fig:210407_dm_structure}
\end{figure*}

\section{\frb\ Galactic Scattering Prediction (NE2025)}\label{app:NE2025}
We compared the measured scattering properties of \frb\ with Galactic expectations using the NE2025 electron density model~\citep[][]{Ocker_2026}. The line-of-sight refractive-index structure parameter ($C_n^2$) profile, which quantifies the strength of electron-density turbulence, peaks at a characteristic distance $D_{\rm MW}^{\rm peak}$ (the location of strongest turbulence). For our sightline ($l=249.5^\circ$, $b=-65.7^\circ$; Section~\ref{sec:Detection and Localisation}), NE2025 gives $D_{\rm MW}^{\rm peak}\simeq0.285$\,kpc. For an extended medium, the effective single-screen position $D_{\rm MW,eff}$ is the geometric-weighted centroid of the $C_n^2$ distribution:
\begin{equation}
    D_{\rm MW,eff}=\frac{\int_0^s C_n^2(s)\,s\left(1-\frac{s}{D_{\rm FRB}}\right)\,ds}{\int_0^s C_n^2(s)\,ds},
\end{equation}
where $s$ is the distance from the observer and $D_{\rm FRB}$ is the angular diameter distance between the source and the observer, computed from the measured redshift $z = 0.2106$ using Planck 2018 cosmology~\citep{2020A&A...641A...6P} with \texttt{astropy} \citep[][]{2022ApJ...935..167A}, giving $D_{\rm FRB} = 7.3176 \times 10^5$\,kpc. This yields $D_{\rm MW,eff}\simeq 0.9392$\,kpc.

At 919.5~MHz, NE2025 predicts $\tau^{\rm NE2025} = 7.67 \times 10^{-5}$\,ms and $\Delta \nu_{\rm d}^{\rm NE2025} = 2.41$\,MHz (assuming $V_{\rm ISS}=100$\,km\,s$^{-1}$), from \citet[][Equations~9 and~10]{cordes2002}. These satisfy the single-screen Fourier uncertainty relation $2\pi\tau_{\rm sc}\Delta\nu_{\rm d}\approx 1$. By contrast, our measured values give $2\pi\tau_{\rm sc}\Delta\nu_{\rm d}\sim 5600$, indicating that $\tau_{\rm sc}$ and $\Delta\nu_{\rm d}$ cannot arise from a single thin screen.

\subsection{Host galaxy screen size}\label{app:host screen}
NE2025 additionally provides the extragalactic angular-broadening-weighted scattering measure, $\mathrm{SM}_{\theta,x}$ \citep[their Equation~6]{cordes2002}, from which the Milky Way screen's scattering-disk size can be estimated independently of the resolved two-screen formalism (Section~\ref{sec: low m}). Using the extragalactic (plane-wave) form of the angular-broadening relation \citep[their Equation~8]{cordes2002},
\begin{equation}\label{eq:theta-xgal}
    \theta_{\rm FWHM} = 128\,\mathrm{SM}_{\theta,x}^{3/5}\,\nu^{-11/5}\quad \mathrm{mas},
\end{equation}
with $\nu$ in GHz and $\mathrm{SM}_{\theta,x}$ in kpc\,m$^{-20/3}$, NE2025 gives $\mathrm{SM}_{\theta,x} = 2.5504\times10^{-4}$\,kpc\,m$^{-20/3}$ at 919.5\,MHz, so that
\begin{equation}
    \theta_{\rm FWHM} \approx 1.075~\mathrm{mas}.
\end{equation}
Converting to the 2$\sigma$ width convention of \citet{pradeepe.t.2025a} (their Equation~3.6), $\theta_{L}=(2/\sqrt{8\ln2})\,\theta_{\rm FWHM}$, gives
\begin{equation}
    \theta_{L,\rm MW} \approx 0.913~\mathrm{mas}.
\end{equation}
Adopting $D_{\rm MW,eff}\simeq0.9392$\,kpc, the corresponding physical screen size 
\citep[][their Equation~3.9]{pradeepe.t.2025a} is
\begin{equation}
    L_{\rm MW} = 4\,\theta_{L,\rm MW}\,D_{\rm MW,eff} \approx 3.43~\mathrm{AU}.
\end{equation}
This estimate is derived independently of the measured $m_g$ and $\Delta\nu_{d,\rm MW}$, and is consistent with the value obtained from the measured scintillation properties in Section~\ref{sec:two screen} ($L_{\rm MW}\approx3.04$\,AU).

\section{Faraday Rotation and Scattering of a Time-Varying PA}\label{app:RM scatter}
Here we consider whether the \RM{s} of two Faraday-rotating screens remain additive if they are separated by an operation that mixes time and frequency dependence, such as scattering.

We can write the intrinsic linear polarisation of a burst in complex polar form as
\begin{equation}
    \mathcal{L}(t) = L e^{2i\psi_0(t)},
\end{equation}
where $L=\sqrt{Q^2+U^2}$ is the linearly polarised intensity and $\psi_0(t)$ is the intrinsic, time-varying PA. The first screen applies \RM$_1$,
\begin{equation}
    \mathcal{L}_1(t,\nu) = L e^{2i[\psi_0(t)+{\rm RM}_1\lambda^2]}.
\end{equation}

We then convolve this with a scattering kernel, $h(t;\nu)$,
\begin{align}
    \mathcal{L}_2(t,\nu) &= \int h(t-t';\nu)\mathcal{L}_1(t',\nu)\,{\rm d}t' \\
                         &= L e^{2i{\rm RM}_1\lambda^2} \int h(t-t';\nu)e^{2i\psi_0(t')}\,{\rm d}t' \\
                         &= L e^{2i{\rm RM}_1\lambda^2}S(t,\nu),
\end{align}
where
\begin{equation}
    S(t,\nu) \equiv \int h(t-t';\nu)e^{2i\psi_0(t')}\,{\rm d}t'.
\end{equation}

The PA after scattering is therefore
\begin{align}
\psi_2(t,\nu) &= \frac{1}{2}\arg[\mathcal{L}_2(t,\nu)] \\
              &= \frac{1}{2}\arg[S(t,\nu)] + {\rm RM}_1\lambda^2 \\
              &= \psi_{\rm scat}(t,\nu) + {\rm RM}_1\lambda^2.
\end{align}

Although the intrinsic PA, $\psi_0(t)$, has no frequency dependence, the scattering convolution can introduce frequency dependence into $\psi_{\rm scat}(t,\nu)$ because the scattering kernel depends on frequency and mixes emission from different intrinsic PAs. Consequently, the apparent \RM\ can become time dependent across the burst.

The final screen applies \RM$_2$,
\begin{align}
    \mathcal{L}_3(t,\nu) &= \mathcal{L}_2(t,\nu)e^{2i{\rm RM}_2\lambda^2} \\
                         &= L S(t,\nu)e^{2i({\rm RM}_1+{\rm RM}_2)\lambda^2}.
\end{align}

The final PA is therefore
\begin{equation}
    \psi_3(t,\nu) = \psi_{\rm scat}(t,\nu) + ({\rm RM}_1+{\rm RM}_2)\lambda^2.
\end{equation}

Thus, at the level of the polarisation signal, the two \RM{s} remain additive, and their order relative to the scattering operation does not affect the total \RM. However, the scattering operation can introduce an additional frequency- and time-dependent contribution to the PA. Defining a local apparent \RM\ as
\begin{align}
    {\rm RM}_{\rm app}(t) &\equiv \frac{\partial\psi_3(t,\nu)} {\partial\lambda^2},
\end{align}
we obtain
\begin{align}
    {\rm RM}_{\rm app}(t) &= \frac{\partial\psi_{\rm scat}(t,\nu)} {\partial\lambda^2} + {\rm RM}_1+{\rm RM}_2.
\end{align}
Thus, the physical \RM{s} remain additive, while the apparent \RM\ contains an additional scattering-induced term,
\begin{equation}
    {\rm RM}_{\rm scat}(t) \equiv \frac{\partial\psi_{\rm scat}(t,\nu)} {\partial\lambda^2},
\end{equation}
which depends on the intrinsic PA evolution, $\psi_0(t)$, and the frequency-dependent scattering kernel. This term may bias an \RM\ estimate obtained from the observed PA evolution.

\end{document}